\documentclass{aa}  
\usepackage{comment}

\usepackage{xcolor}

\usepackage{graphicx}
\usepackage{txfonts}

\usepackage{arydshln} 
\usepackage{pifont}
\usepackage{placeins}

\usepackage{natbib}
\bibpunct{(}{)}{;}{a}{}{,} 

\usepackage[colorlinks=true, urlcolor=blue, colorlinks=true, citecolor=blue, linkcolor=blue]{hyperref}
\begin{document} 
      
   \title{The properties of magnetised cold filaments in a cool-core galaxy cluster}
    
   \author{M. Fournier
          \inst{1}, P. Grete\inst{1}, M. Brüggen\inst{1}, F. W. Glines\inst{2,3} and B. W. O'Shea\inst{2,3,4}
         }

   \institute{Universität Hamburg, Hamburger Sternwarte, Gojenbergsweg 112, 21029 Hamburg, Germany \\
              \email{martin.fournier@hs.uni-hamburg.de}
        \and
    Department of Computational Mathematics, Science, and Engineering, Michigan State University, East Lansing, MI 48824, USA
         \and
    Department of Physics and Astronomy, Michigan State University, East Lansing, MI 48824, USA
        \and
    Facility for Rare Isotope Beams, Michigan State University, East Lansing, MI 48824, USA
}
   \date{Received \today}

  \abstract
   {Filaments of cold gas ($T\leq 10^{4}$ K) are found in the inner regions of many cool-core clusters. These structures are thought to play a major role in the regulation of feedback from active galactic nuclei (AGNs).}
   {We study the morphology of the filaments, their formation, and their impact on the propagation of the outflowing AGN jets.}
   {We present a set of GPU-accelerated 3D magnetohydrodynamic simulations of an idealised Perseus-like cluster using the performance portable code {\scshape{AthenaPK}}. We include radiative cooling and a self-regulated AGN feedback model that redistributes accreted material through kinetic, thermal, and magnetic feedback.}
   {We confirm that magnetic fields play an important role in both the formation and evolution of the cold material. These suppress the formation of massive cold discs and favour magnetically supported filaments over clumpy structures. Achieving resolutions of $25-50$ pc, we find that filaments are not monolithic as they contain numerous and complex magnetically supported sub-structures. We find that the mass distribution of these clumps follows a $\mathrm{d}N/\mathrm{d}M \propto M^{-1.6}$ power-law for all investigated filaments. Studying the evolution of individual filaments, we find that their formation pathways can be diverse. We find examples of filaments forming through a combination of gas uplifting and condensation, as well as systems of purely infalling clumps condensing out of the intracluster medium. The density contrast between the cold gas and the outflowing hot material leads to recurring deflections of the jets, favouring inflation of bubbles.}
   {Filaments in cool-core clusters are clumpy and contain numerous sub-structures, resulting from a complex interplay between magnetic fields, thermal instability, and jet-cloud interaction. Frequent deflections of the AGN outflows suppress jet collimation and favour the formation of large X-ray bubbles, and smaller off-axis cavities.}
    
   \keywords{galaxies: clusters: intracluster medium – galaxies: jets – galaxies: clusters: general – methods: numerical – magnetohydrodynamics
               }

   \maketitle
%

\setlength{\parindent}{0cm}

\section{Introduction}

Galaxy clusters are massive ($M \sim 10^{14} - 10^{15} \ \rm{M}_\odot$) structures resulting from the gravitational collapse of the highest over-densities in the early Universe and their subsequent accretion of matter. Dark matter accounts for roughly 85\% of the cluster halo mass, with baryons accounting for only about 15\%. Over 90\% of the galaxy cluster’s baryonic mass is in the form of a hot ($10^7-10^8$ K) diffuse plasma sitting in hydrostatic equilibrium in the dark matter halo. \citep{Andreon_2010,Gonzalez_2013,Laganda_2013}.\\

The intracluster medium (ICM) is expected to cool down radiatively through bremsstrahlung, recombination, and collisional processes \citep{Boehringer_1989}. A third to a half of all clusters, usually designated as cool-core (CC) clusters, have a large central over-density with an associated cooling time as small as two orders of magnitudes below the Hubble time \citep{Peres_1998}. This suggests that in the absence of heating sources, cooling gas should flow towards the cluster's centre, with mass rates of up to $10^{3} \, \rm{M}_\odot \ \rm{yr}^{-1}$ \citep{Fabian_1994,White_1997}. The absence of specific spectral signatures of such cooling flows from X-ray surveys or comparably large star formation rates in the clusters' central galaxies leads to the conclusion that such gas flows are absent in most clusters \citep{Peterson_2001} and that a heating source must be counteracting the radiative cooling. Observations of CC clusters shed light on the presence of large cavities apparently inflated by active galactic nucleus (AGN) jets \citep{Fabian_2006,Birzan_2004}. While it is generally accepted that AGNs are the main heating source balancing radiative cooling in galaxy clusters, the way this energy is transferred and distributed in the surrounding ICM remains an open question \citep{Bohringer_2010}. \\

Observations of various clusters suggest that the gas inside the inner tens of kiloparsecs of clusters is multi-phase, with temperatures ranging from $10$ to $10^{9}$ K \citep[e.g.]{Allen_2000,Bonamente_2001}. H$\alpha$-emitting nebulae of cold, elongated filaments have been widely studied, and are thought to be ubiquitous in CC clusters \citep{McDonald_2012,Olivares_2019}. Multi-wavelength surveys of CC clusters show that the filaments have typical lengths ranging from a few kiloparsecs to tens of kiloparsecs, and thicknesses down to $\sim 50$ pc \citep{Fabian_2008,Fabian_2016}, and that they are typically found below or at the edge of the X-ray cavities \citep{Salome_2006,Tremblay_2018,Olivares_2019}, suggesting a tight connection between AGN feedback and the formation of the filaments. It is commonly assumed that thermal instability (TI) is a key process contributing to the formation of multi-phase gas \citep{McCourt_2012,Voit_2017}. Turbulence, induced either by AGN jets or by environmental effects, drives the formation of over-densities in the ICM. Due to the dependence of radiative cooling on density, these over-densities can grow non-linearly by drawing hot material out of the background ICM, resulting in structures of molecular and ionised gas with electron densities of $\gtrsim
 100 \, \rm{cm}^{-3}$ \citep{McDonald_2012}. These structures are then expected to decouple from the ICM and ‘rain’ towards the centre of the cluster, eventually fueling the supermassive black hole (SMBH) in a chaotic manner \citep{Gaspari_2013}. It is also suspected that the AGN jets may be able to uplift cold material present in the inner kiloparsecs of the clusters up to altitudes of tens of kiloparsecs \citep{Russell_2017,Tremblay_2018}. This uplifted material could act as a seed for condensation and act as a precursor to the formation of massive, infalling filaments. \\

 Various simulation set-ups have been employed to test this model. Hydrodynamic and magnetohydrodynamic (MHD) simulations of stratified or periodic boxes \citep{McCourt_2012,Valentini_2015,Mohapatra_2022} and idealised CC clusters or massive elliptical galaxies \citep{Li_2014,Beckmann_2019,Qiu_2019,Wang_2021,Guo_2023,Ehlert_2023,Guo_2024} have been able to reproduce some of the observed properties of cold filaments. In \citet{Li_2014}, the interaction of AGN outflows with the ICM leads to the formation of cold clouds along the jet axis. However, the resulting structures are very clumpy and lack the coherent and elongated morphologies observed in surveys of CC clusters. Other simulations with coarser spatial resolutions (typically a few hundred parsecs) have been performed to include jet re-orientation \citep{Beckmann_2019}, radiative transfer \citep{Qiu_2019}, magnetic fields \citep{Wang_2021,Ehlert_2023}, cosmic rays \citep{Wang_2020,Beckmann_2022}, and anisotropic thermal conduction \citep{Beckmann_2022}. Magnetic fields are thought to play a significant role in the formation of elongated threads of cold gas, suppressing the formation of cold clumps as well as massive cold discs that are common in hydrodynamic set-ups \citep{Li_2014,Qiu_2019}. However, MHD simulations of idealised CC clusters with spatial resolutions resolving the inner structure of the filaments (i.e. down to scales of a few tens of parsecs) are still lacking, and consequences for AGN feedback remain largely unconstrained. In this paper, we present a set of high-resolution, 3D hydrodynamic and MHD simulations of an idealised Perseus-like galaxy cluster with a turbulent, radiatively cooling ICM and self-regulated AGN feedback. Our GPU-accelerated simulations were performed on a static, nested mesh that was refined down to a maximum spatial resolution of 24.4 pc.\\

In Sect.~\ref{Setup}, we present the numerical set-up of our simulations and describe the parameters used for each of our runs. In Sect.~\ref{ClusterEvol}, we discuss the overall evolution of the ICM properties over the entire duration of our simulations. In Sect.~\ref{Properties}, we study the morphology and main physical properties of the cold gas filaments found in our high--resolution MHD run. In Sect.~\ref{Formation}, we take a closer look at the formation of several filaments and discuss the relative contribution of gas uplifting and condensation. In Sect.~\ref{Impact}, we assess the impact of cold gas on the morphology of AGN outflows, before discussing our results and assessing the importance of missing physical process in our simulation in Sect. \ref{Discussion} and concluding in Sect.~\ref{Conclusion}.

\section{Simulations and method}
\label{Setup}

\subsection{Magnetohydrodynamics with radiative cooling}

We used the performance-portable {\scshape{AthenaPK}} code based on the block-structured refinement framework {\scshape{Parthenon}} \citep{Parthenon}. The fluxes were computed using Harten-Lax-van Leer-contact (HLLC) \citep{HLLC} and Harten-Lax-van Leer discontinuity (HLLD) \citep{HLLD} solvers for our hydrodynamic and MHD runs, respectively. Overall, a second-order accurate method was used based on piecewise linear reconstruction, a predictor-corrector van-Leer-type integrator, and, in the MHD case, a generalised Lagrange multiplier-based divergence cleaning approach \citep{Dedner2002}. The basic system of equations is given by

\begin{equation}
    \begin{aligned}
        &\frac{\partial \rho}{\partial t} + \nabla \cdot [\rho \mathbf{u}] = 0, \\
        &\frac{\partial \rho \mathbf{u}}{\partial t} + \nabla \cdot \left[ \rho \mathbf{u} \otimes \mathbf{u} + P_{\text{tot}} \mathbb{I} - \mathbf{B}\otimes\mathbf{B} \right] = 0, \\
        &\frac{\partial e_{\text{tot}}}{\partial t} + \nabla \cdot  \left[ \mathbf{u}(e_{\text{tot}} + P_{\text{tot}}) - \mathbf{u}\cdot\mathbf{B})\mathbf{B} \right] + \mathcal{L} (\rho,T) = 0, \\
        &\frac{\partial \mathbf{B}}{\partial t} - \nabla \times [\mathbf{u} \times \mathbf{B}] = 0. \\
    \end{aligned}
\label{MHD}
\end{equation}

Here, $\rho$ is the gas density, $\mathbf{u}$ is its velocity field, $\mathbf{B}$ is the magnetic field vector, and $P_{\text{tot}}$ is the total pressure, including the thermal pressure, $P=(1-\gamma)e$, and the magnetic field pressure, $P_{\text{mag}} = B^2/2$. The gas is assumed to be non-relativistic, and thus we take $\gamma=5/3$. $e$ is the gas internal energy density and $e_{\text{tot}}$ designates the total energy density, $e_{\text{tot}}=e + 0.5 \rho u^2 + B^2/2$.

$\mathcal{L}(\rho,T)$ is a function modelling the optically thin cooling processes taking place in the gas component.
We include radiative cooling of gas down to $10^{4.2}$ K using the exact integration scheme introduced in \citet{Townsend} and the cooling table from \citet{Schure}, assuming half solar metallicity \citep{Fabian_2011}. A temperature floor was set to $T_{\rm{floor}}=10^4$ K.
Finally, given the non-relativistic set-up of our simulations, we also limited the temperature
to $T<5\times10^9$\,K, and gas and Alfv\'en velocities to $0.05$\,c in the central region
of the simulation ($r < 20$\,kpc).

\subsection{Grid structure}

We adopted a set-up akin to that employed by \citet{Wang_2021} and Grete et al. (in prep.). A root grid of $512^3$ cells covering a volume of (800 kpc)$^{3}$ was statically refined in its central region. The width of cells at refinement level $\ell$ is given by

\begin{equation}
    \Delta x_{\ell} = \frac{L_{\rm{box}}}{512 \times 2^{\ell}}.
\end{equation}

In our highest-resolution runs, we used $\ell = 6$ refinement levels. The three finest refinement levels, with cell sizes of 24.4 pc, 48.8 pc, and 97.6 pc, cover the inner 6.4 kpc, 12.6 kpc, and 25 kpc of the box, respectively. Such grid geometry allows us to simulate $\sim80\%$ of the total cold gas mass formed with cells of 48.8 pc at most in our highest-resolution runs (see Appendix \ref{spatialresolutionstudy}).

\subsection{Initial conditions}
\label{sect:initialconditions}

In all our simulations, we impose a constant gravitational potential. The dark matter halo of the cluster was modelled using a Navarro-Frenk-White (NFW) profile \citep{Navarro_1997}:

\begin{equation}
    \begin{aligned}
        g_{\text{NFW}}(r) &= \frac{\mathcal{G}}{r^2} \frac{M_{\text{NFW}} \ \left[ \ln\left(1 + \frac{r}{R_{\text{NFW}}} \right) - \frac{r}{r + R_{\text{NFW}}} \right]}{\ln(1+c_{\text{NFW}}) - \frac{c_{\text{NFW}}}{1 + c_{\text{NFW}}}}, 
    \end{aligned}
\end{equation}
where $\mathcal{G}$ is the gravitational constant, $M_{\text{NFW}}$ is the NFW profile virial mass, $c_{\text{NFW}}$ its concentration parameter, and $R_{\text{NFW}}$ is a scale radius function of the virial mass and the concentration parameter, defined by the following relations:

\begin{equation}
    \begin{aligned}
    &R_{\rm{NFW}} = \left ( \frac{M_{\rm{NFW}}}{ 4 \pi \rho_{\rm{NFW}} \left [ \ln{\left ( 1 + c_{\rm{NFW}} \right )} - c_{\rm{NFW}}/\left(1 + c_{\rm{NFW}} \right ) \right ] }\right )^{1/3},\\
&\rho_{\rm{NFW}} = \frac{200 \, H_0^2}{8\pi\mathcal{G}} \frac{c_{\rm{NFW}}^3}{\ln{\left ( 1 + c_{\rm{NFW}} \right )} - c_{\rm{NFW}}/\left(1 + c_{\rm{NFW}} \right )}.
 \end{aligned}
\end{equation}

Secondly, a Hernquist potential was used to model NGC1275, the brightest cluster galaxy (BCG), with $M_{\text{BCG}}$ and $R_{\text{BCG}}$ its mass and radius \citep{Hernquist_1990}:

\begin{equation}
    \begin{aligned}        
        g_{\text{BCG}}(r) &= \frac{\mathcal{G}M_{\text{BCG}}}{R_{\text{BCG}}^2} \left(1+\frac{r}{R_{\text{BCG}}}\right)^{-2} .
    \end{aligned}
\end{equation}

Finally, we added a central point mass modelling the effect of the SMBH of mass $M_{\rm{BH}}$. Density, pressure, and temperature profiles were then calculated from this gravitational field so that the ICM obeys a hydrostatic equilibrium. We assume an entropy profile, $K(r)$, defined by

\begin{equation}
    K \equiv \frac{k_B T(r)}{n_e^{2/3}(r)},
\end{equation}
where $k_B$ is the Boltzmann constant, $r$ is the radial distance to the centre of the cluster, $T(r)$ is the temperature at a given distance, $r$, and $n_e(r)$ is the electron density. The profile follows the empirical power law introduced in the ACCEPT database \citep{ACCEPT_2009}:

\begin{equation}
    K(r) = K_0 + K_{100} \ \left( \frac{r}{100 \text{ kpc}} \right)^{\alpha_K}.
\end{equation}

In order to close the system of equations defining the hydrostatic equilibrium, we fixed the electron density of the cluster to a defined value, $n_e(r_{\rm{ref}})$, at a defined radius, $r_{\rm{ref}}=10$ kpc.

\begin{table}[h!]
\caption{Values adopted for the various initial condition parameters in our simulation runs}
\centering
\begin{tabular}{lccc}\hline\hline
Parameter       & Value                                                  & Unit                        & \multicolumn{1}{r}{Reference} \\ \hline
$H_0$          & $75.1$              & $\rm{km} \, \rm{Mpc}^{-1} \, \rm{s}^{-1}$              & \textbf{a}\\
$M_{\rm{NFW}}$ & $6.6\times 10^{14}$ & $\rm{M}_\odot$              & \textbf{b}\\
$c_{\rm{NFW}}$ & 5.0                                                    & -                           &                               \\
$R_{\rm{BCG}}$ & $10$                                                   & $\rm{kpc}$                  &  \textbf{c}                             \\
$M_{\rm{BCG}}$ & $2.4\times10^{11}$                                     & $\rm{M}_\odot$              &  \textbf{d}                             \\
$M_{\rm{BH}}$  & $4\times 10^{8}$                                       & $\rm{M}_\odot$              &  \textbf{e}                             \\
$K_0$          & $10.0$                                                 & $\text{ keV} \ \text{cm}^2$ &  \textbf{f}                             \\
$K_{100}$      & $150.0$                                                & $\text{ keV} \ \text{cm}^2$ &  \textbf{f}                             \\
$\alpha_K$     & $1.1$                                                & -                           & \textbf{f}                                \\
$n_e(r_{\rm{ref}})$ & $0.05$                                            & $\rm{cm}^{-3}$              &  \textbf{g}                             \\ \hline
\end{tabular}
\tablebib{\textbf{a}) \citet{Hubble_2021}, \textbf{b}) \citet{Simionescu_2011}, \textbf{c}) \citet{Vaucouleurs_1991}, \textbf{d}) \citet{Mathews_2006}, \textbf{e}) \citet{Park_2017}, \textbf{f}) \citet{ACCEPT_2009}, \textbf{e} \citet{Churazov_2003}.}
\label{table:parametersNFW}
\end{table}

In order to break the symmetry of the system, we added velocity perturbations in each run. 40 wave vectors with length scales between 12.5\,kpc and 50\,kpc were sampled using a random seed and their relative contribution was computed from an inverse parabolic spectrum. We took a peak wavelength of 25 kpc and a velocity dispersion of 75 km s$^{-1}$. We used a similar spectrum to initialise a perturbed magnetic field, whose average local amplitude follows the density profile

\begin{equation}
    \langle B(r) \rangle \propto \rho(r)^{2/3}.
\end{equation}

The strength of the initial magnetic fields was chosen such that its average intensity in the inner few kiloparsecs of the box is $\sim 10 \, \mu\rm{G}$ \citep{Taylor_2006}. The same random seed was used in each of our runs so that the perturbed fields were identical from one simulation run to another. The values adopted for each of these parameters are summarised in Table \ref{table:parametersNFW}.

\subsection{Active galactic nuclei feedback}
\label{agnfeedback}

We modelled the effect of the central SMBH on its environment by including AGN feedback through three different channels; namely, thermal, kinetic, and magnetic feedback.

\subsubsection{Cold gas accretion}
\label{ColdAccretion}

Since the ICM is thermally unstable, turbulence is expected to trigger the formation of cold clouds, which decouple from the surrounding hot gas and fall chaotically towards the centre of the cluster. Accordingly, we followed the chaotic cold accretion model developed in \citet{ChaoticColdAccretion} and later implemented in cluster-scale simulations \citep{Li_2014,Meece2017}. We used an AGN triggering model based on the accretion of cold material. We defined the accretion region as a sphere of radius $R_{\rm{acc}}=500$ pc, and selected the gas with a temperature below a threshold value of $T_{\rm{cold,acc}}=5\times 10^4 \,\rm{K}$ as the accretion material. The accretion rate at each time step was then derived by summing mass over all cells containing gas beneath $T_{\rm{cold,acc}}$ and within the accretion region:
\begin{equation}
    \dot{M}_{\rm{acc}} = \sum_{r_i < R_{\rm{acc}}} \frac{\rho(\mathbf{\hat{r}}_i) \ dV(\mathbf{\hat{r}}_i)}{t_{\rm{acc}}} \ \delta(T(\mathbf{\hat{r}}_i) < T_{\rm{cold,acc}}),
\end{equation}
where $\rho(\mathbf{\hat{r}}_i)$ is the gas density in cell $i$, and $d\rm{V}$ is that cell's volume. Cold gas in these cells was depleted over the accretion timescale $t_{\rm{acc}}=5$ Myr. This accretion rate was then used to compute the AGN feedback power, given by

\begin{equation}
    \label{AGN_power}
    \dot{E}_{\text{AGN}} = \eta \ {\dot{M}}_{\rm{acc}} c^2,
\end{equation}
where $\eta = 10^{-2}$ is the cold mass triggering efficiency \citep{Gaspari_2011}. The energy injected in each feedback channel is then a fraction of $\dot{E}_{\text{AGN}}$, given by

\begin{equation}
    \dot{E}_{X,\text{AGN}} = f_X \ \dot{E}_{\text{AGN}},
\end{equation}
where $X$ designates the feedback mode (i.e. $f_T$, $f_K$, and $f_M$ for thermal, kinetic, and magnetic feedback, respectively, see Table~\ref{table:parameters}).

\subsubsection{Thermal feedback}

We modelled the radiative feedback from the AGN by depositing a portion of the AGN feedback power, $f_T \cdot \dot{E}_{\rm{AGN}}$, through thermal energy within a sphere of radius $R_T=500 \,\rm{pc}$. The corresponding power is related to the AGN power (Eq. \ref{AGN_power}) through

\begin{equation}
    \dot{e}_T(r \leq R_T) = \frac{3}{4} \ \frac{f_T \ \dot{E}_{\rm{AGN}}}{\pi R_T^3}.
\end{equation}
Mass was also injected within the same region of radius $R_T$:

\begin{equation}
    \dot{\rho}_T(r \leq R_T) = \frac{3}{4} \ \frac{f_T \ \dot{M}_{\rm{acc}}}{\pi R_T^3}.
\end{equation}

\subsubsection{Kinetic feedback}

We modelled the jets by redistributing material in two cylinders of radius $r_{\rm{jet}}=4\,\Delta x_{\rm{min}}$, thickness $h_{\rm{jet}}=2\,\Delta x_{\rm{min}}$, and with an offset of $L_{\rm{jet}}=4\,\Delta x_{\rm{min}}$ with respect to the centre of the box. We positioned these cylinders in a static orientation along a fixed axis. The velocity of the jet was determined by
\begin{equation}
    v_{\rm{jet}} = \sqrt{2\eta} \ c,
\end{equation}
with $c=3\times 10^8\,\rm{km}\,\rm{s}^{-1}$ the speed of light. The density of the deposited material is
\begin{equation}
    \rho_{\rm{jet}} = \frac{(1-\eta) f_K\dot{M}_{\rm{acc}}}{2\pi r_{\rm{jet}}^2 h_{\rm{jet}}}.
\end{equation}

In order to track the jet, material as passive scalar was evolved whose concentration was set to unity within the jet launching regions.

\begin{figure*}[h!]
    \includegraphics[width=\textwidth]{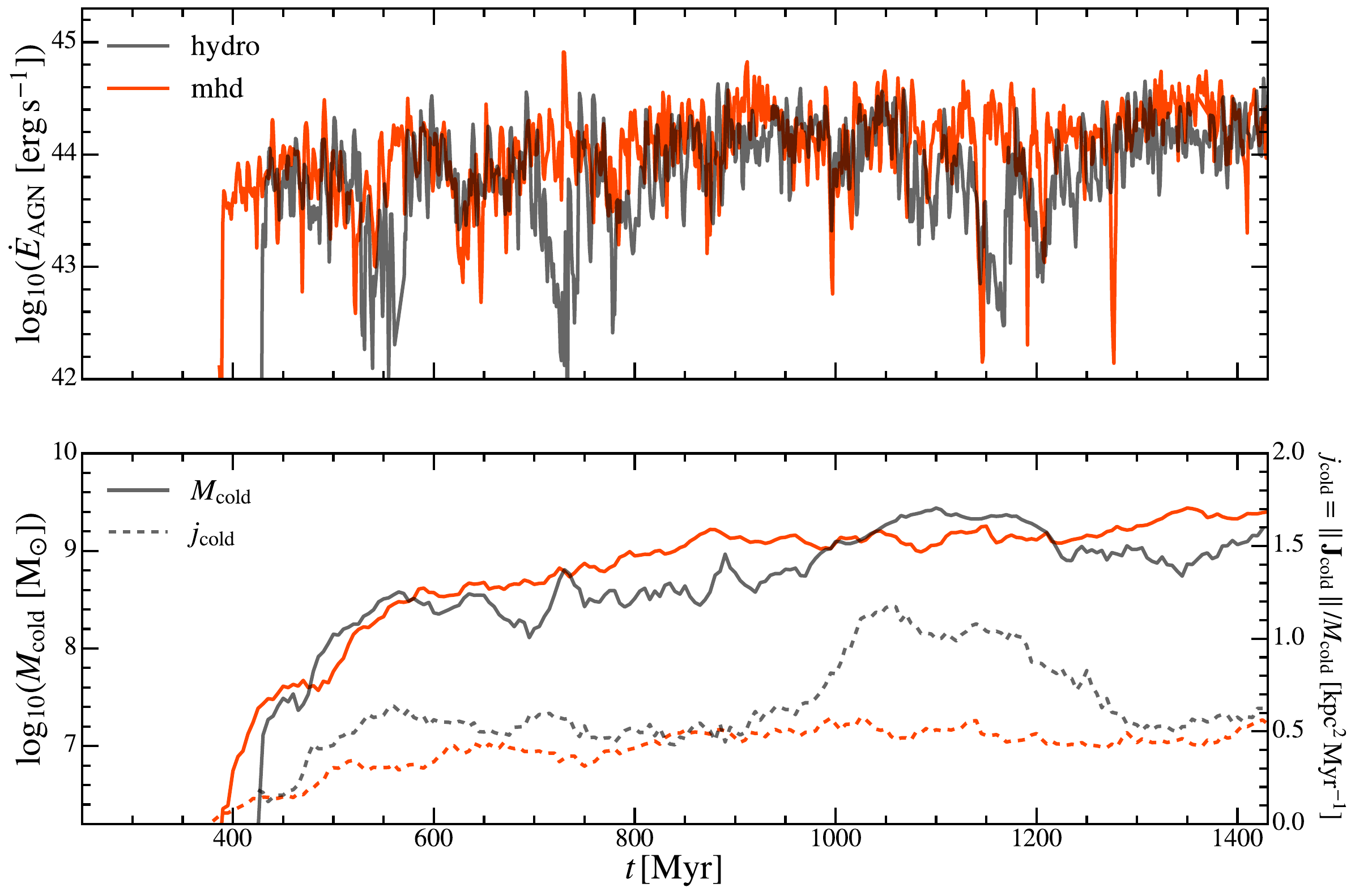}
    \caption{Evolution of the AGN power (top panel) as a function of time for our highest-resolution hydrodynamic (dark blue) and MHD (light blue) runs. The bottom panel presents the evolution of the total cold gas mass formed (solid line), as well as its associated specific angular momentum (dashed line). The average AGN power is broadly constant from one set-up to an other, but shows weaker fluctuations in MHD set-up. While similar quantities of cold gas are formed in each run, its overall specific angular momentum is found to be larger in the hydrodynamic case due to the formation of a massive disc, consistent with previous work \citep{Li_2014,Qiu_2019,Ehlert_2023}.}
    \label{fig:PAGN}
\end{figure*}

\subsubsection{Magnetic feedback}

The importance of magnetic fields in the acceleration and collimation of AGN jets has been emphasised by numerous theoretical and observational studies. General relativistic MHD simulations have shown that these are dynamically important in both the launching and propagation of the jets \citep{Liska_2020,Lalakos_2024}, while Faraday rotation measurements in AGN jets leads to strengths ranging from $10^2 \,\rm{G}$ in launching regions down to typical ICM strengths of a few $\mu$G \citep{Kino_2015,Park_2019,Gomez_2022}. \\ 

We included magnetic feedback in our MHD runs by depositing purely toroidal magnetic field loops in all cells at the same height as the kinetic feedback cylinders (i.e. within $L_{\rm{jet}} \leq \vert z \vert \leq L_{\rm{jet}} + h_{\rm{jet}}$), following:

\begin{equation}
    \mathcal{B}_{\theta}(r) = 2 \mathcal{B}_0 \ \left(\frac{r}{L_B}\right) \ \exp \left[ - \left(\frac{r}{L_B}\right)^2 \right],
\end{equation}
in cylindrical coordinates.
$L_B=2.5\,\Delta x_{\rm{min}}$ is the characteristic length scale of the toroidal field set so that the injected fields
are mostly confined to the launching region and $\mathcal{B}_0$ was dynamically adjusted so that the resulting
injected magnetic energy matches the magnetic fraction of the target AGN feedback.

\subsection{Stellar feedback}

\subsubsection{Type Ia supernova feedback}

Type Ia supernova (SNIa) feedback was included, following the method described in
\citet{Prasad2020}. The energy density and mass density injected into a cell at a distance, $r$, from the cluster centre is

\begin{equation}
    \begin{aligned}
        &\dot{e}_{\rm{SNIa}} = \Gamma_{\rm{SNIa}} \ E_{\rm{SNIa}} \ \rho_{\rm{BCG}}(r), \\
        &\dot{\rho}_{\rm{SNIa}} = \alpha \ \rho_{\rm{BCG}}(r),
    \end{aligned}
\end{equation}
where $\Gamma_{\rm{SNIa}} = 3 \times 10^{-14} \, \rm{yr}^{-1} \ \rm{M}_{\odot}^{-1}$ is the SNIa rate per unit mass, $E_{\rm{SNIa}} = 10^{51} \rm{erg}$ is the amount of energy released per supernova event, and $\alpha = 10^{-19} \ \rm{s}^{-1}$ is the mass injection rate.

\subsubsection{Stellar formation}

In the absence of separate star particles, we also instantaneously converted gas that
would end up in stars into thermal energy.
More specifically, a fraction of the gas above a number density threshold of $n > 50 \, \rm{cm}^{-3}$, a temperature below $T<2\times10^4$\,K, and within a radius of $R_\mathrm{acc} < r < 25$\,kpc was locally converted into heat with an efficiency of $5\times10^{-6}$.

\subsection{Simulation parameters}

\begin{table}[h!]
\caption{Overview of the parameters used in our runs.}
\label{table:parameters}
\centering
\begin{tabular}{lccccc}\hline\hline
Name        & \begin{tabular}[c]{@{}c@{}}Max. res.\\ {[}pc{]}\end{tabular} & $f_K$ & $f_T$ & $f_M$ & \begin{tabular}[c]{@{}c@{}}$\langle B(r=0) \rangle$\\ {[}$\mu$G{]}\end{tabular} \\ \hline
hydro       & 24.4                                                           & 0.75  & 0.25  & -     & -                                                                               \\ \hline
mhd         & 24.4                                                           & 0.74  & 0.25  & 0.01  & 10 \\ \hline
\end{tabular}
\tablefoot{Columns indicating (from left to right): simulation name; width of cells in the maximally refined region of the box; feedback fractions for kinetic, thermal, and magnetic energy; average initial magnetic field amplitude at the centre of the box.}
\end{table}

We ran two high-resolution simulations, whose parameters are summarised in Table \ref{table:parameters}. In addition to our high-resolution run with a minimum cell width of 24.4 pc, we performed three hydrodynamic and three MHD simulations with fewer refinement levels in order to study the impact of numerical resolution on our results. The dependence of our results on spatial resolution is discussed in Appendix \ref{spatialresolutionstudy}.
\section{Cluster evolution}
\label{ClusterEvol}

\begin{figure}[h!]
    \centering
    \resizebox{\hsize}{!}
    {\includegraphics{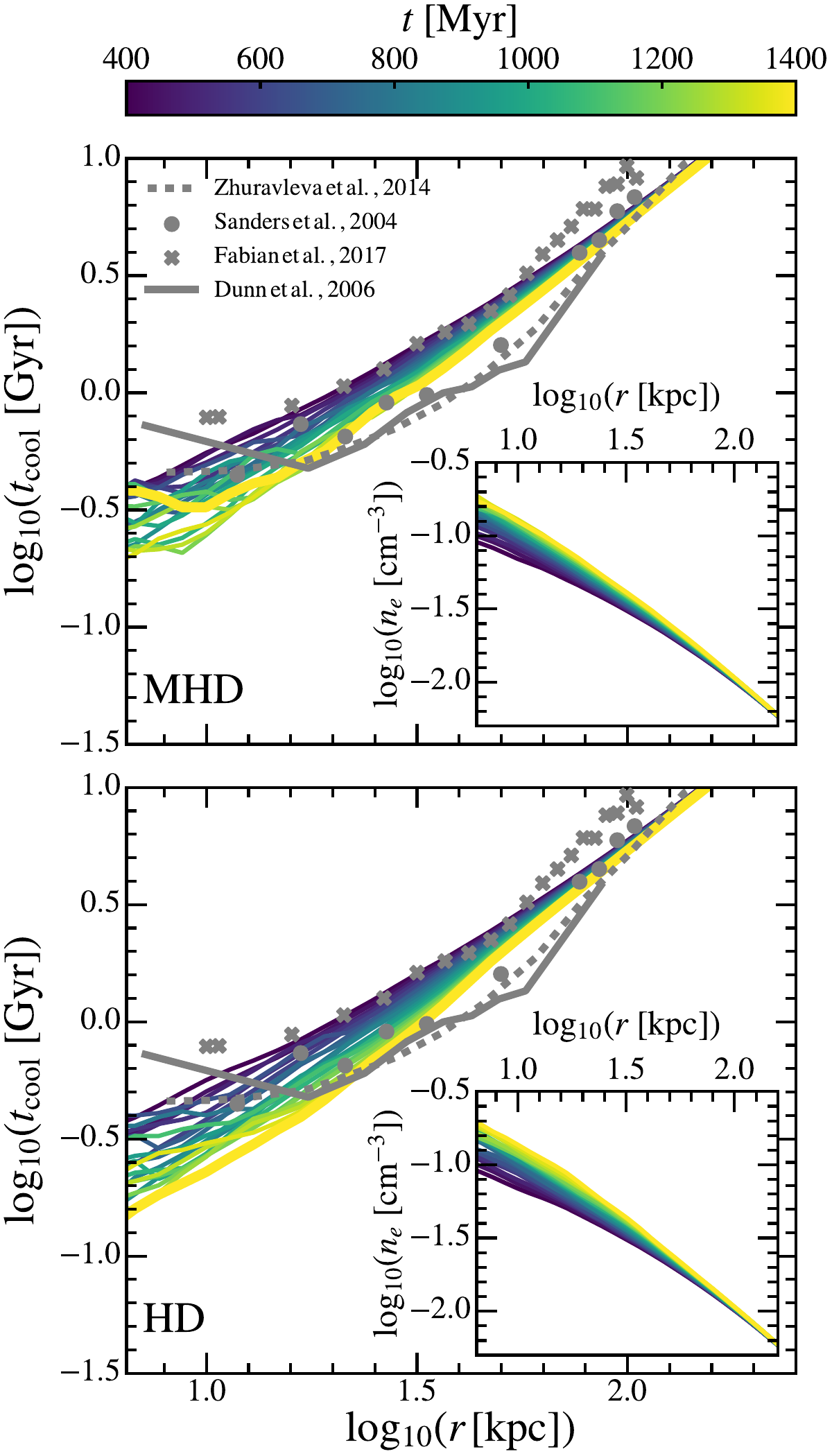}}
    \caption{Evolution of the radial profiles for cooling time and electron density evaluated at various times in our highest-resolution MHD (top) and hydrodynamic (bottom) runs for the hot ICM gas (i.e. $T\geq 10^5$ K). Observational constraints for the Perseus cluster are also indicated for cooling time for comparison \citep{Sanders_2004,Dunn_2006,Zhuravleva_2014,Fabian_2017}.}
    \label{fig:ProfileVsTime}
\end{figure}

Because our AGN triggering mode is only sensitive to cold gas, the SMBH is inactive during the first $\sim$400 Myr of our simulation. During that time, the gas properties evolve due to radiative cooling and the initial velocity turbulence. Over-densities build up in the centre of the box and end up forming cold clumps after roughly 400 Myr. In Fig.~\ref{fig:PAGN}, we present the evolution of the AGN feedback power (top panel), total cold gas mass (bottom panel, solid line), and its associated specific angular momentum (dashed line) as a function of time for our high-resolution MHD and HD simulations. The average AGN power is roughly constant across  runs and is of the order of $10^{44} \ \rm{erg} \, \rm{s}^{-1}$. Phases of low AGN power down to $\sim 10^{42} \ \rm{erg} \, \rm{s}^{-1}$ lasting for a few tens of millions of years are visible in the hydrodynamic run, with a typical period of 100~--~200 Myr. The overall feedback power remains much steadier in the MHD run, with weaker fluctuations that last for shorter periods. The total cold gas mass formed throughout our simulations converges to typically $\sim 10^9 \ \rm{M}_\odot$ after $\rm{1}$ Gyr, consistent with observational constraints \citep{Salome_2006,Fabian_2008} and previous simulations \citep{Beckmann_2019,Li_2014,Ehlert_2023}. We find that the specific angular momentum of the cold gas is less than in our MHD set-up. A massive cold disc quickly forms in our hydrodynamic run, while we find no such structure in our MHD runs. Such behaviour is in agreement with previous simulation work \citep{Wang_2020,Ehlert_2023}. In Fig.~\ref{fig:ProfileVsTime}, we show the cooling time and electron density radial profile of the hot ICM at various times in our MHD and hydrodynamic simulations, from the start of the first AGN phase at $t \sim 400$ Myr to the end of our runs. Observational constraints for the cooling time profile is also presented by the grey lines and points \citep{Sanders_2004,Dunn_2006,Zhuravleva_2014,Fabian_2017}. We find no major differences between our hydrodynamic and MHD set-ups regarding the overall properties of the ICM. The cooling profiles remain broadly stable over time, with values of $2-3 \, \times \, 10^{8} \, \rm{yr}$ in the inner 10 kpc, consistent with observations. The electron density increases in both our runs up to $\sim10^{-1} \, \rm{cm}^{-3}$. At late times, the MHD runs show an elevated cooling time.

\section{Properties of the cold gas}
\label{Properties}

\begin{figure*}[h!]
    \includegraphics[width=\textwidth]{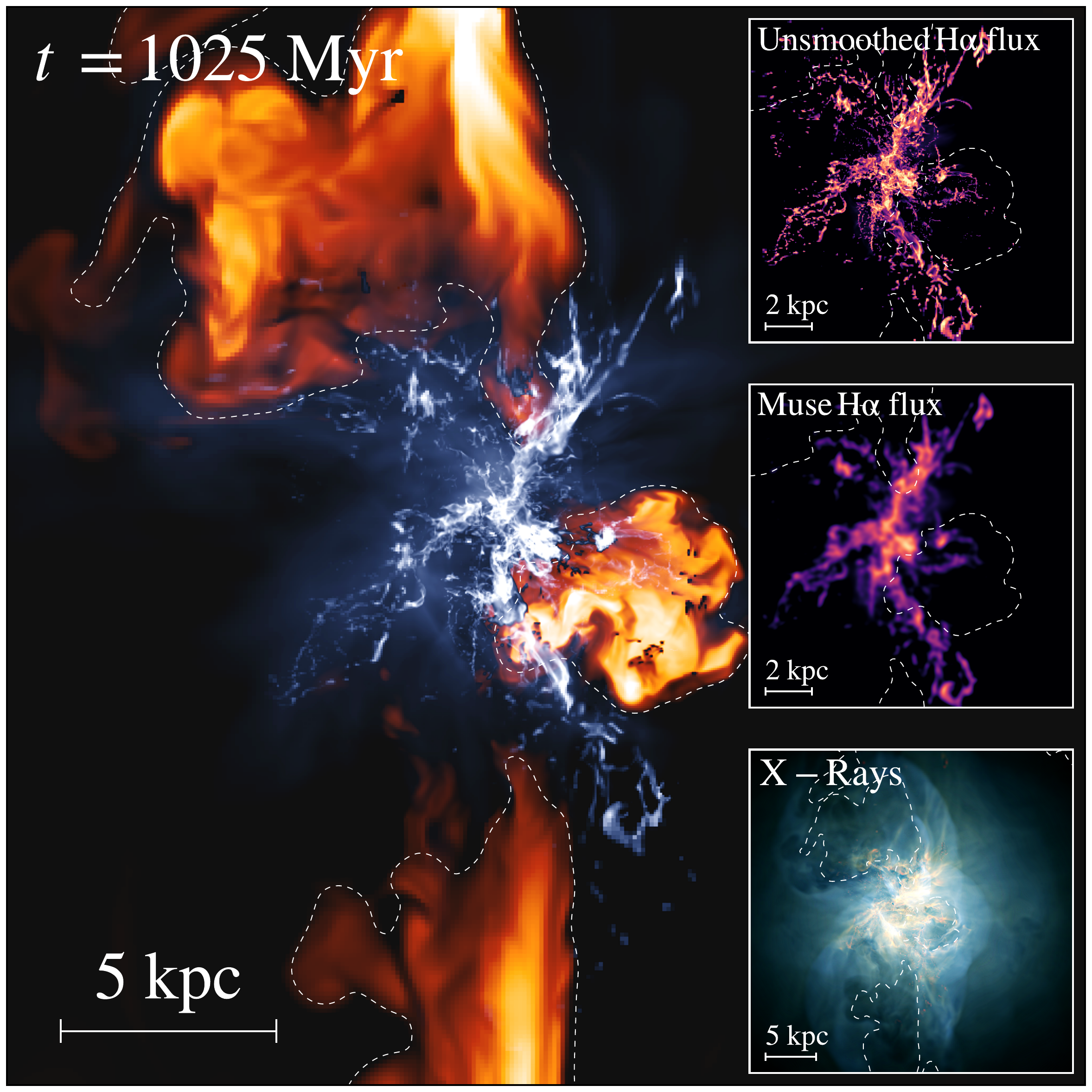}
    \caption{Composite image showing the density-weighted projection of the magnetic field strength (blue) and the temperature (orange) for the inner 25 kpc of our high-resolution MHD run at $t=1025$ Myr. Mock H$\alpha$ and X-ray images are also shown in the lower right of the image. The top H$\alpha$ image is untouched, while the centre right one has been smoothed with a Gaussian filter to match the angular resolution of the Muse instrument, assuming our cluster is at the same distance as the Perseus cluster, and is comparable to Fig. 2 in \citet{Olivares_2019}. The X-ray image is a superposition of three filters; namely, the 0.3–1.2 keV (red), 1.2–2 keV (green), and 2–7 keV (blue) bands, and is comparable to Fig. 1 from \citet{Fabian_2006}. The dashed white line is a contour map from the temperature field, showing the position of the outflowing bubbles.}
    \label{fig:Halpha}
\end{figure*}

\subsection{Cluster visualisation}

To begin our exploration of the cold gas properties in our simulated cluster, we present a series of visualisations from our high-resolution MHD simulation in Fig.~\ref{fig:Halpha}. The main image shows the magnetic field strength (blue), overlaid with a projection of the temperature field (orange). Elongated structures of gas with magnetic field strengths of the order of  $\sim 100 \, \mu$G are visibly interacting with the outflowing AGN jets, deflecting them and inflating several cavities. In the top and centre right of the figure, we present mock H$\alpha$ images of these cold structures. These have been produced by integration along the line of sight of the H$\alpha$ emission rate, $j_{H\alpha}$, from \citet{Dong_2011}:

\begin{equation}
    j_{H\alpha} = 2.83 \times 10^{-26} T_4^{-0.942-0.031\ln{(T_4)}} n_e n(H^+) \, \rm{erg} \, \rm{cm}^3 \, \rm{s}^{-1} \, \rm{sr}^{-1},
\end{equation}

where $T_4 = T / 10^4 \, \rm{K}$ and $T$, $n_e$, and $n(H^+)$ are the temperature, electron density, and ionised hydrogen density of each gas cell in the simulated box, respectively. The second H$\alpha$ flux has been smoothed with a Gaussian filter to match the angular resolution of the MUSE instrument, assuming that our simulated cluster is at the same distance from the observer as the Perseus cluster, and is comparable to Fig. 2 from \citet{Olivares_2019}. Our simulations show clumpy internal sub-structure inside the filaments that is unresolved by MUSE. The image in the lower right is a composite X-ray mock image imitating Fig.~1 from \citet{Fabian_2006}, and traces the outflowing bubbles carved in the ICM by the AGN jets. The red, green, and blue components represent the 0.3–1.2 keV, 1.2–2 keV, and 2–7 keV bands, respectively.

\subsection{Morphology}
\label{section:morphology}

Previous research has shown that magnetic fields affect the overall morphology of the cold gas. While high-resolution hydrodynamic simulations can produce cold gas in quantities consistent with observations, the morphology of the structures is usually very clumpy \citep{Li_2014} and lacks a coherent, filamentary morphology. Magnetic fields have been found in both cluster and cloud-crushing set-ups to inhibit the formation of clumps and massive discs, and to enhance the formation of elongated structures \citep{Wang_2020,Ehlert_2023,Das_2024}. \\

\begin{figure}[h!]
    \includegraphics[width=0.5\textwidth]{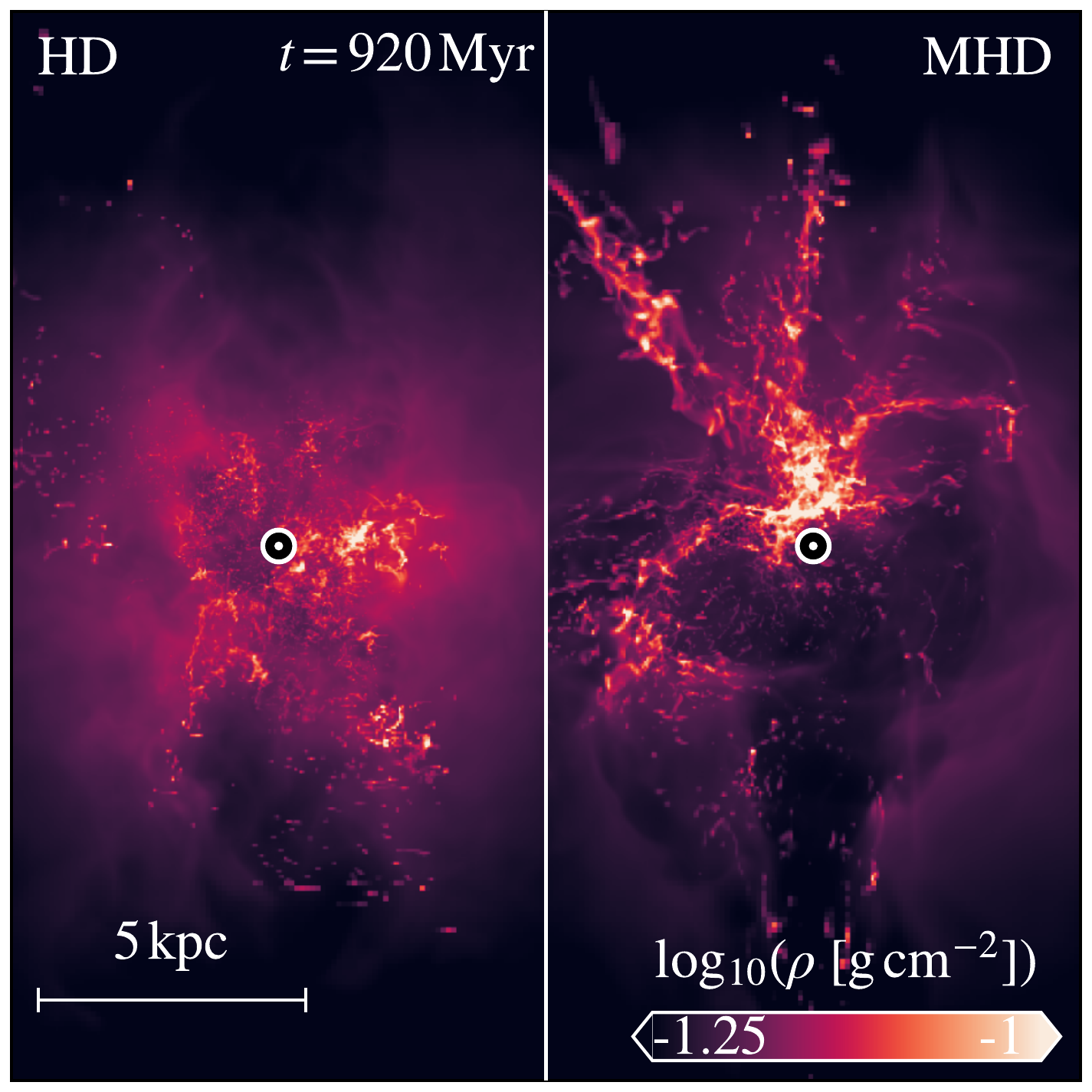}
    \caption{Gas density projection of the inner 20 kpc of our simulated cluster at $t=920$ Myr for our hydrodynamic (left panel) and MHD runs (right panel). The gas mostly gathers into tiny cloudlets of a few tens of parsecs, while magnetic fields promote the formation of elongated filaments.}
    \label{fig:Comparison}
\end{figure}

In Fig.~\ref{fig:Comparison}, we present a gas density projection of the inner 20 kpc in our two high-resolution simulations. The left panel shows the hydrodynamic run and the right panel the MHD run. In agreement with previous work \citep[e.g.]{Ehlert_2023}, we find that the magnetic fields have an important impact on the morphology of the cold gas. In the hydrodynamic case, the cold gas condenses into thousands of cloudlets typically consisting of just 1--10 cells. In our MHD set-up, the gas gathers along extended filamentary structures, preferentially infalling towards the centre of the cluster. Each of these structures is dynamic, reaches maximum elongations up to $\sim 15$ kpc, and contains up to $\sim 10^{8.5} \, \rm{M}_\odot$ of cold gas. \\

The morphologies observed in our hydrodynamic run are hard to reconcile with observations. Optical and radio surveys have shown that cold gas in CC clusters is concentrated in filaments and that massive cold discs are rare \citep{Olivares_2019,Tremblay_2018,Russell_2017}. In the following, and unless stated otherwise, we thus focus on the results from our high-resolution MHD run. \\

\begin{figure}[h!]
    \centering
    \resizebox{\hsize}{!}
    {\includegraphics{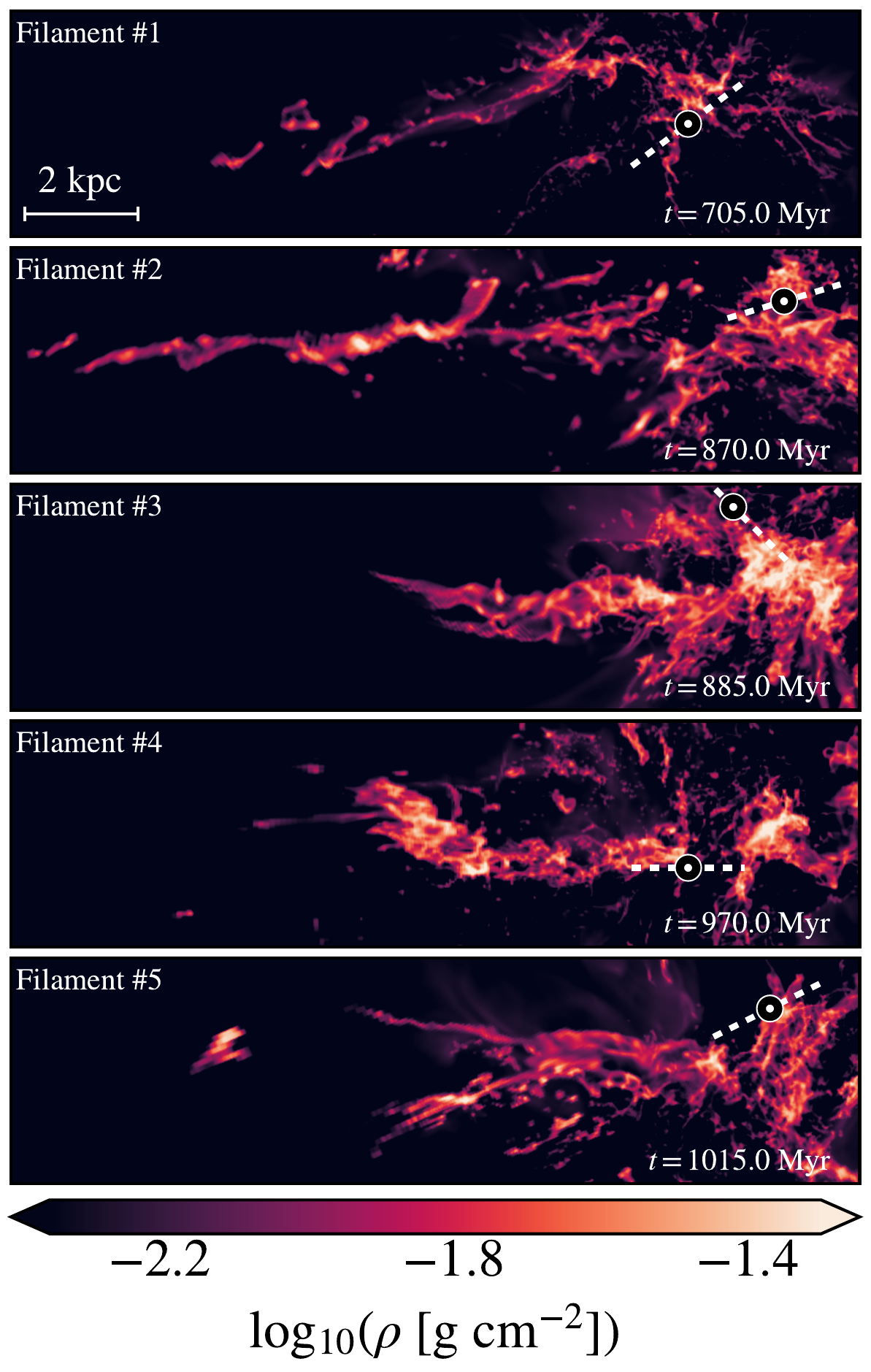}}
    \caption{Projection of the gas density for five different filaments in our fiducial MHD run. Each projection was performed at the time when each filament reached its maximum elongation. The projected position of the SMBH and the projected orientation of the jets is indicated by the round marker and the dotted line, respectively.}
    \label{fig:FilamentsIndex}
\end{figure}

\begin{figure*}[h!]
    \includegraphics[width=\textwidth]{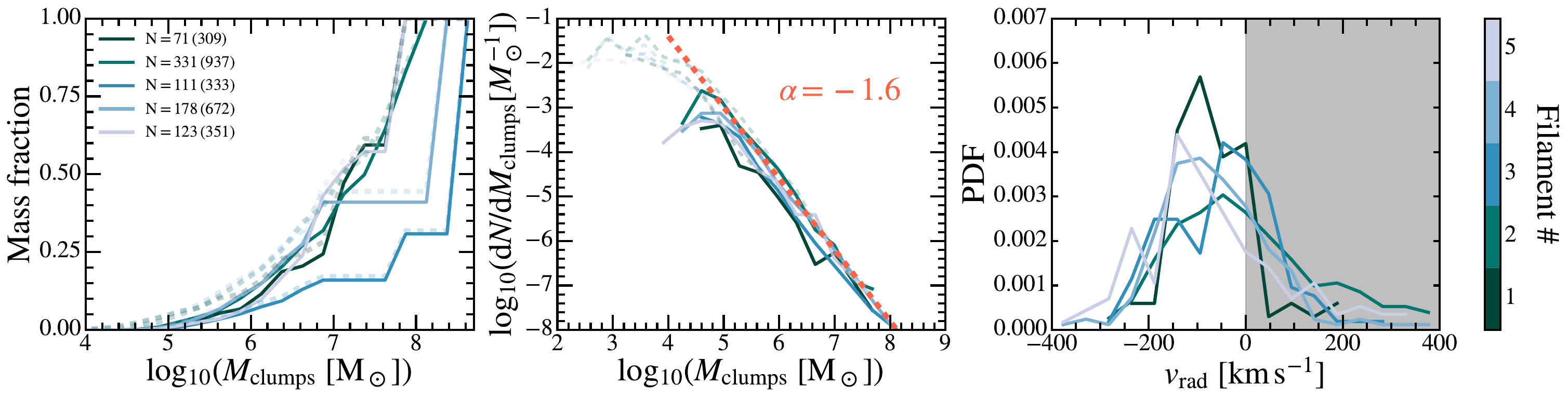}
    \caption{Main statistical properties of the clump sample obtained from the five filaments presented in Fig.~\ref{fig:FilamentsIndex}. The left panel shows the cumulative mass fraction of the clumps as a function of their mass. The distribution of the clumps for the five filaments as a function of their mass and radial velocity are presented in the central and right panel, respectively. The number of clumps for each filament is indicated in the legend of the left panel. The first number indicates the number of clumps included in our statistical analysis, while the number in parentheses indicates the total clumps population with no constraint on the minimum number of cells per clump. The associated clump mass distribution with no constraint on the minimum number of cell is also indicated by the dashed bright blue lines. The dashed orange line indicates a $d\rm{N}/d\rm{M}$ power-law distribution of index $\alpha=-1.6$ for comparison. The shaded grey area in the top right plot corresponds to the region of positive radial velocity (i.e. clumps flowing out from the cluster centre).}
    \label{fig:FilamentStatistics}
\end{figure*}

Among the cold structures observed over the course of our fiducial MHD simulation, we identify five filaments that lend themselves to a morphological study because their relative isolation helps identify their boundaries. In Fig.~\ref{fig:FilamentsIndex}, we present the gas density projections of these five distinct filaments at the time they reach their maximum elongation. These are referred to by their index, which is indicated in the top left of each panel. The typical filament ranges between 5 and 15 kpc in length, and its morphology can be complex. In particular, all filaments exhibit numerous inner structures of cold clumps and sub-filaments of various thicknesses spanning between $\sim 800$ pc and the resolution limit of the simulation. To study quantitatively the distribution of these sub-structures in and around these five filaments, we used a clump-finding algorithm integrated in the visualisation and data analysis package {\scshape{yt}} \citep{Turk_2011,yt4}. The algorithm operates by first defining a minimum density threshold. It then employs a contouring method to recursively construct a hierarchical structure of clumps and sub-clumps. In our set-up, we used a minimum density threshold of $n_{\rm{clump,min}} = 1 \, \rm{cm}^{-3}$ and a step factor of $s=4$ corresponding to the multiplication factor applied to the minimum density threshold at each iteration of the algorithm. The impact of the clump-finding algorithm is discussed in Appendix~\ref{impactclumpfinder}. For each system, we limited the search for dense structures to all gas cells contained in a cylinder of radius $r_{\rm{cyl}}=2$ kpc manually aligned with each filament's main axis. We also excluded all clumps with fewer than ten cells and with a mean temperature above $T_{\rm{clump,max}}=10^5$ K to exclude possible artefacts from the hot ICM. Several cell volume-weighted quantities were calculated for each filament by summing over all gas cells belonging to a given clump. In particular, we derived the magnetic field strength, the ratio of the thermal pressure to magnetic pressure, $\beta$ (later on referred as plasma, $\beta$), and the radial velocity of each clump belonging to each of the five filaments. \\

In Fig.~\ref{fig:FilamentStatistics}, we present the cumulative mass fraction of the clumps belonging to each filament, as well as the probability distribution function (PDF) of the clumps as a function of their gas mass (central panel) and radial velocity (right panel). $N$ in the left panel indicates the number of clumps identified in each filament included in our analysis; that is, each one containing 10 or more cells. The corresponding PDFs are represented by solid blue lines. The number in parentheses corresponds to the total number of clumps with no constraint on the minimum number of cell for each clump. The associated PDFs are also represented by dashed blue lines for comparison. 

As is visible in the left panel, a large fraction of the mass of each filament is carried by massive clumps. In particular, the cumulative mass of all clumps with individual masses of $M_{\rm{clump}} \leq~10^{7} \ \rm{M}_\odot$ contributes $\sim$ 25 \% of the overall mass of each filament, while 50 \% of each filament's mass is carried by the few clumps with masses of $M_{\rm{clump}} \geq 10^{7.5} \ \rm{M}_\odot$. The number of such massive clumps is low and around 1--5 for each filament system.

The overall shape of the mass distributions (central panel) is remarkably constant across filaments and seems to follow a power law between $10^{5} \, \rm{M}_\odot$ and $10^7-10^8 \, \rm{M}_\odot$. The corresponding index of the clump mass distribution, $\rm{d}N/\rm{d}\rm{M}$, is typically of $\alpha = -1.6$ (see dashed orange line for comparison). This power-law index is close to the value of -2 reported in numerous other studies, including turbulent box simulations \citep{Gronke_2022,Fieding_2023,Das_2024,Tan_2024}, idealized cluster simulations \citep{Yuan_2014b}, and cosmological simulations \citep{GIBLE,Augustin_2025}. \\

The left and central panels support a picture in which the filaments are clumpy structures, largely dominated in number by light clumps with masses as low as $10^4 \, \rm{M}_\odot$, and largely dominated in mass by the contribution of only a few massive clumps with individual masses above $10^7 - 10^8 \, \rm{M}_\odot$. We emphasise that most of the lightest clumps remain spatially unresolved, and that increasing spatial resolutions might impact the shape of the mass distribution, especially at its lower mass end. Indeed, the position of the peak of the distribution is shifted to lower masses when including the excluded clumps with fewer then ten cells, suggesting that the lower mass end of the clump statistics has not yet converged. This is expected and consistent with results from 2D and 3D cloud-crushing simulations that conclude that the fragmentation of cooling gas produces cloudlets with typical radii reaching down to sub-parsec scales, well below our resolution limit of 25 pc \citep{McCourt_2018,Jennings_2023}. 

In the right panel of Fig.~\ref{fig:FilamentStatistics}, we show the distribution of clumps as a function of radial velocity. As is visible, the bulk motion of the filaments is preferably infalling, with average velocities of between --100 and --200 km s$^{-1}$. All five filaments contain clumps characterised by a positive radial velocity, up to +400 km s$^{-1}$. These clumps are likely uplifted due to their interaction with the AGN jets, and might play a role in the mass build-up of the filaments (see Sect.~\ref{Formation}).

\begin{figure*}[h!]
    \includegraphics[width=\textwidth]{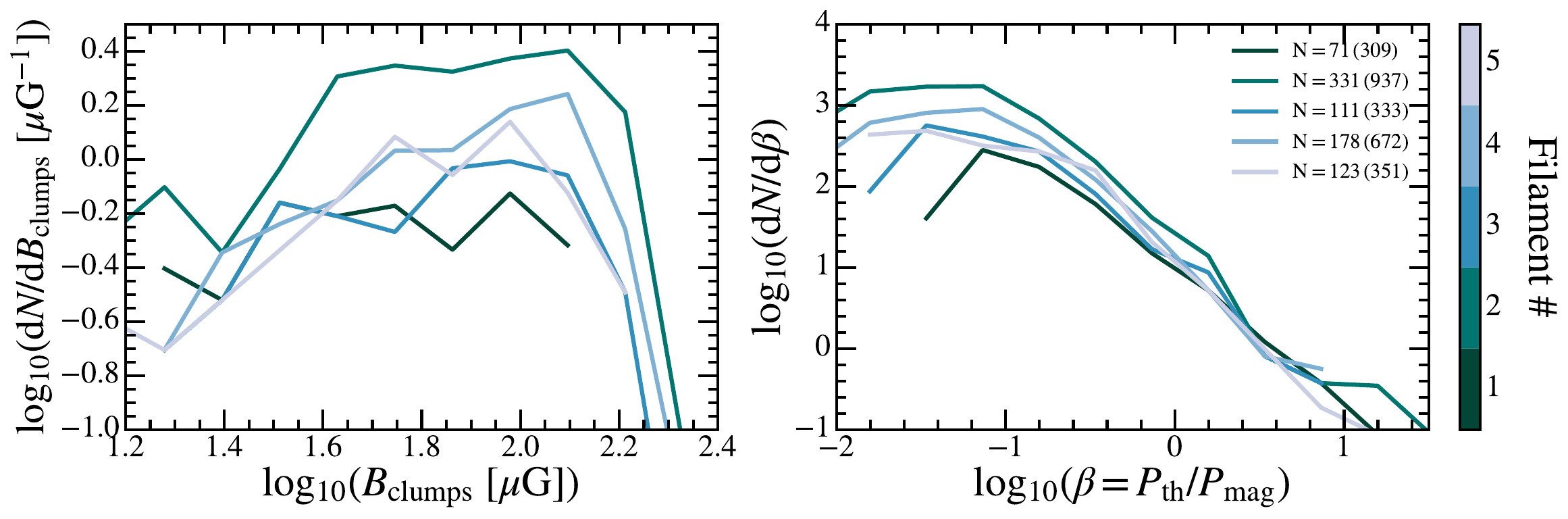}
    \caption{Distribution of the clumps as a function of their volume-weighted magnetic field strength (left panel) and their thermal-to-magnetic pressure ratio, $\beta$ (right panel), for the five filaments presented in Fig.~\ref{fig:FilamentsIndex}.}
    \label{fig:MagneticStat}
\end{figure*}

\subsection{Magnetic support}

In Fig.~\ref{fig:MagneticStat}, we present the distribution of the clumps as a function of their volume-weighted magnetic field strength (left panel) and plasma, $\beta$. The median values of the magnetic field strength in the clumps sample of our five filaments are typically of $\rm{Med}(B_{\rm{clumps}}) \sim 80-100 \ \mu$G. Some of the clumps have a particularly high inner magnetic field strength, ranging up to $\sim200 \ \mu$G. As is shown in the right panel of Fig.~\ref{fig:MagneticStat}, the magnetic pressure is the main source of pressure supporting the filaments against the thermal pressure of the ICM for more than half the clumps. 

\begin{figure*}[h!]
    \includegraphics[width=\textwidth]{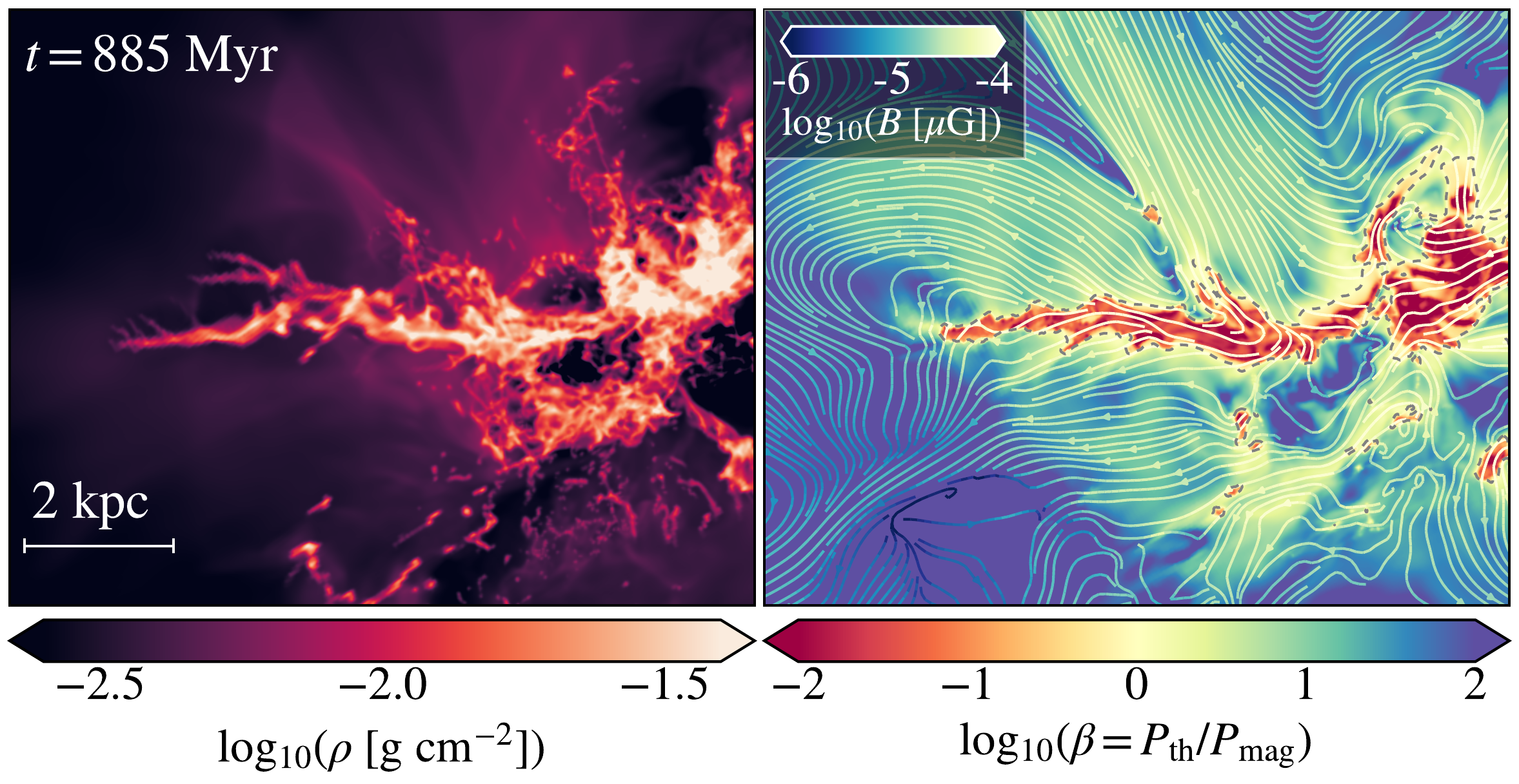}
    \caption{Projection of the gas density (left) and slice of the plasma, $\beta$, overlaid with the magnetic field lines (right) for the filament \#3. The projection depth of the density plot is 4 kpc. The dashed grey line around the filament core represents the contour of equi-partition between magnetic and thermal pressure. (\href{https://www.youtube.com/playlist?list=PLe_DeluPkGwM2tXKomlp8vUiQOXuYhKre}{Movies available})}
    \label{fig:magnetictopology}
\end{figure*}

We find a correlation between the magnetic field lines orientation within the filaments and each filament's main axis. As an example, we show the magnetic field topology of the filament system \#3 in Fig.~\ref{fig:magnetictopology} (see also \href{https://www.youtube.com/playlist?list=PLe_DeluPkGwM2tXKomlp8vUiQOXuYhKre}{associated movies}). The left panel shows the gas column density, highlighting the overall structure of the clump system. The right panel shows a slice of the plasma, $\beta$, in the mid-plane of the system, overlaid with the magnetic field lines. The dashed grey line indicates the contour of equipartition between magnetic and thermal pressure. The magnetic field lines within the regions dominated by magnetic pressure are aligned with the main axis of the filament. This behaviour is also found in the other filaments we have inspected (see also Fig.~\ref{fig:FilamentSequence}). The magnetic field lines outside of the filament are typically uncorrelated with the filament's main axis. In certain regions, the lines are found to flow towards the filament with an orthogonal orientation, and are strongly bent by up to $\sim$130$^{\circ}$ to reach alignment with the filament.

\begin{figure}[h!]
    \centering
    \resizebox{\hsize}{!}
    {\includegraphics{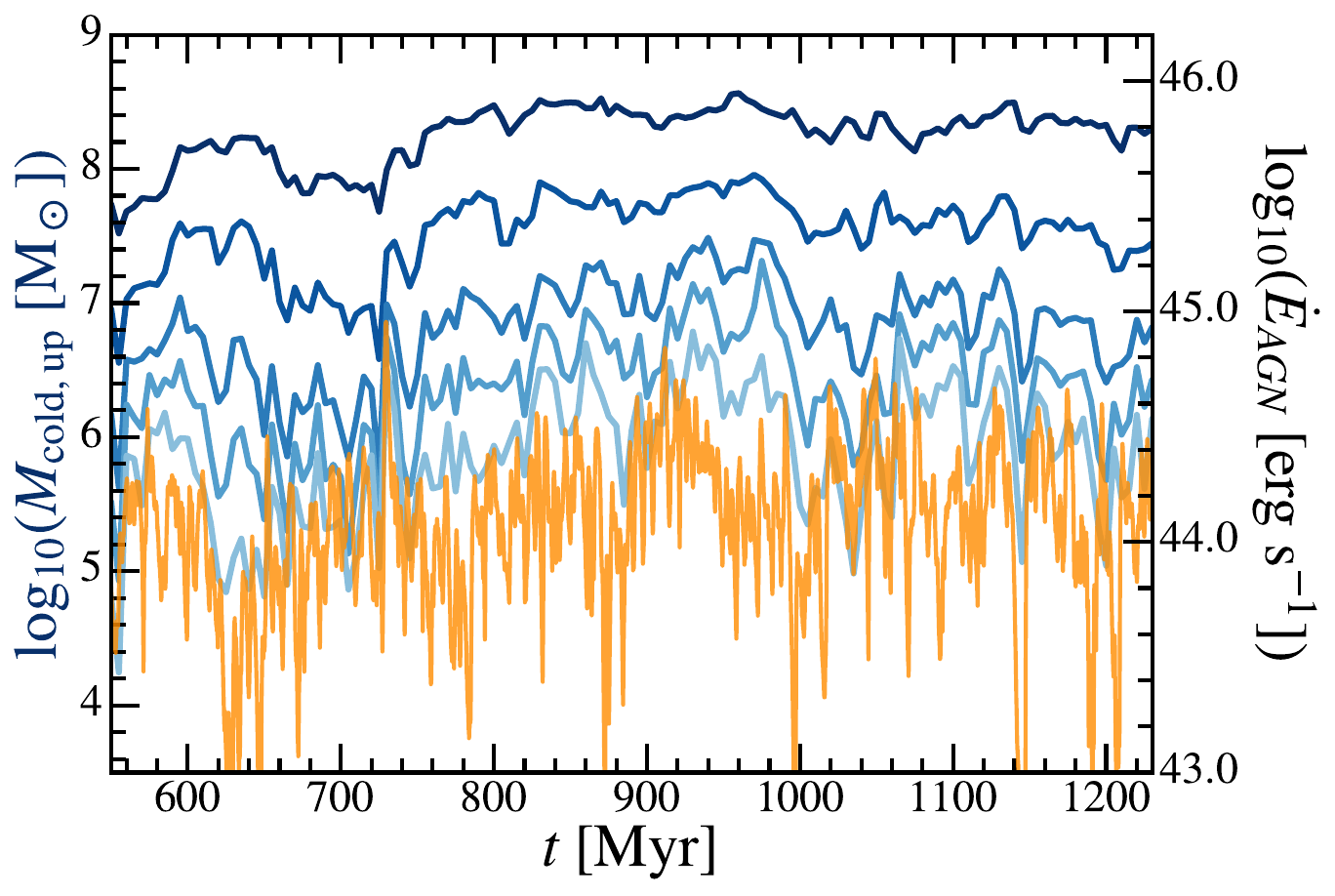}}
    \caption{Total mass of upward-moving cold gas with radial velocities above 0, 150, 300, 400, and 500 km s$^{-1}$ (from dark to bright blue). The solid orange line indicates the evolution of the AGN feedback power.}
    \label{fig:ColdUp}
\end{figure}

\section{Formation of the cold filaments}
\label{Formation}

\subsection{Upward-moving gas and uplifting}

\begin{figure*}[h!]
    \includegraphics[width=\textwidth]{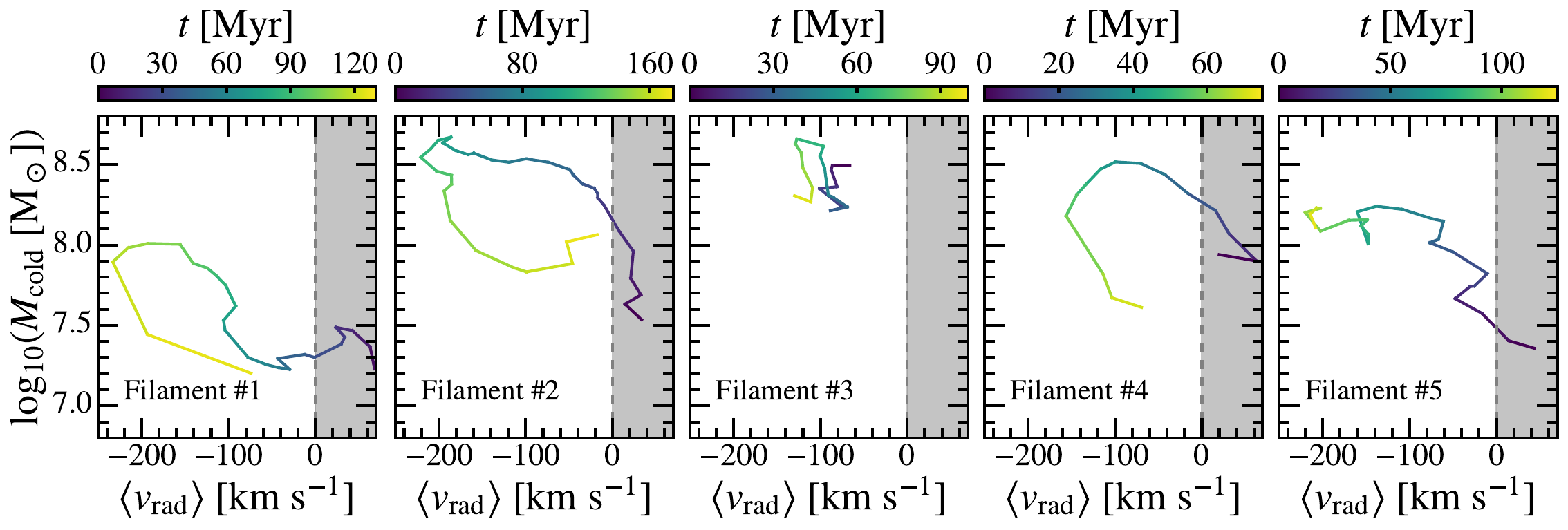}
    \caption{Evolution of the average radial velocity of our five filament systems, as a function of their total cold gas mass budget. The shaded grey area indicates positive radial velocities. For three of our filament systems (namely, \#1, \#2, \#4, and \#5), the bulk of the filament is moving upwards at early times due to its interaction with the AGN outflows. The radial velocity of the system typically reverses $\sim$ 20--40 Myr after the beginning of the positive radial velocity phase. The filaments then continue building up mass while falling back towards the centre of the clusters for $\sim$ 20--100 Myr, before being shredded. The time is re-scaled based on visual inspection such that $t=0$ coincides with the last snapshot before the formation sequence of each filament starts. For filaments \#1, \#2, \#4, and \#5, this corresponds to when the first clumps are visibly starting to move upwards in the direction of what will later on become the filament. For filament \#3, there is no visible upward bulk motion as the filament solely forms out of condensation; thus, $t=0$ is defined here as the last snapshot before infalling clumps visibly gather to form the filament.}
    \label{fig:FilamentsPhasePlot}
\end{figure*}

\begin{figure*}[h!]
    \includegraphics[width=\textwidth]{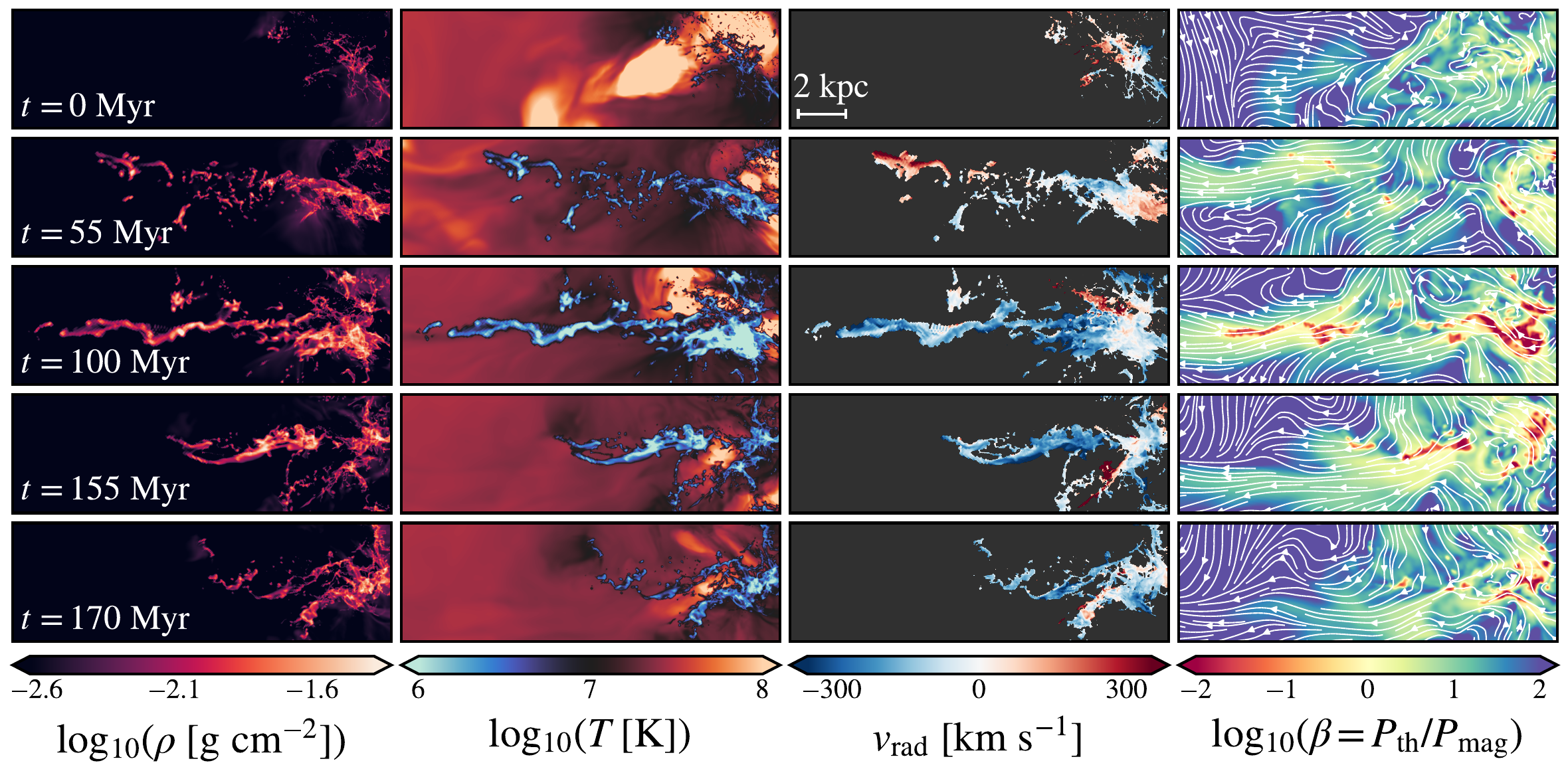}
    \caption{Sequence of images spanning the whole lifetime of filament system \#2 (see Sec.~\ref{section:morphology}). The panels show, from left to right: the gas column density, density-weighted radial velocity, density of jet scalar, and density-weighted thermal-to-magnetic pressure ratio, $\beta$, overlaid with the magnetic field lines. In each case, the plotting region is aligned with the filament's main axis. The SMBH is located to the right of the plot and the jet axis has an angle of $\sim10^{\circ}$ with the main filament axis. Time has been re-scaled such that $t=0$ coincides with the last snapshot before the first clumps starts to move upwards along the subsequent filament's main axis. The times of the snapshots have been selected such that the main steps of the filament evolution are most apparent, and are thus voluntarily unevenly spaced.}
    \label{fig:FilamentSequence}
\end{figure*}

We find evidence of ubiquitous upward-moving gas throughout the entire course of our fiducial MHD simulation. In Fig.~\ref{fig:ColdUp}, we show the evolution of $M_{\rm{cold,up}}$, the total mass of upward-moving cold gas with radial velocities of a certain threshold, defined as

\begin{equation}
    M_{\rm{cold,up}}(v_{\rm{rad}}) = \sum_{i} m_i \, \delta (v_{\rm{rad,i}} \geq v_{\rm{rad}}) \, \delta (T_i \leq 10^5 \, \rm{K}),
\end{equation}

with the sum running over every cell of the box, designated by their index, $i$. The various curves indicate the evolution of $M_{\rm{cold,up}}$ for the following radial velocity thresholds: 0, 150, 300, 400, and 500 km s$^{-1}$ (from dark to bright blue lines). For comparison, the orange curve indicates the evolution of the AGN feedback power, $\dot{E}_{\rm{AGN}}$.

After about 400 Myrs, the total upward-moving gas mass converges to $10^{8.4} \ \rm{M}_\odot$, which is about 15\% of the total cold gas mass (e.g. see Fig.~\ref{fig:PAGN}). There is no obvious correlation between this quantity and the AGN power as the latter flickers on timescales down to less than a million years, while the typical variations in the cold gas mass are smoother and typically of a few million years at least. As the minimum radial velocity threshold is increased (i.e. for brighter blue curves), correlations between peaks of AGN power and peaks of upward-moving cold mass start to be noticeable. In particular, some peaks and drops in AGN power are correlated with peaks and drops in $M_{\rm{cold,up}}$, especially for radial velocity thresholds~$\geq300$~km~s$^{-1}$.\\

In Fig.~\ref{fig:FilamentsPhasePlot}, we show the evolution of the total cold mass as a function of mass-weighted average radial velocity for the five filament systems presented in Fig. \ref{fig:FilamentsIndex}. Four of these filaments (namely, filaments \#1, \#2, \#4, and \#5) exhibit an initial phase where clumps move upwards, with average radial velocities ranging from 20 -- 100 km~s$^{-1}$. Since not all clumps are moving upwards and because the region within which the physical quantities were computed also contains infalling clumps, the average radial velocity is a lower estimate of the overall upward motion of the cold material as individual clumps can move with radial velocities far beyond these values, up to 400 km~s$^{-1}$. For filament systems $\#2$ and $\#4$, the total amount of gas present in the system during the uplifting phase is 30 \% and 50 \% of the maximum mass reached by the filament in its lifetime, respectively. The phase during which the bulk of the filament is characterised by a positive radial velocity lasts for 20 -- 40 Myr in these four filaments, after which the system starts falling back towards the centre of the cluster. Filament \#3 exhibits a different behaviour as no clear evidence of upward-moving cold gas is found prior to the formation of the filament through visually inspection. It is likely that this system forms solely from condensation of the ICM. \\

\subsection{Evolution of the filaments structure}

\begin{figure}[h!]
    \centering
    \resizebox{\hsize}{!}
    {\includegraphics{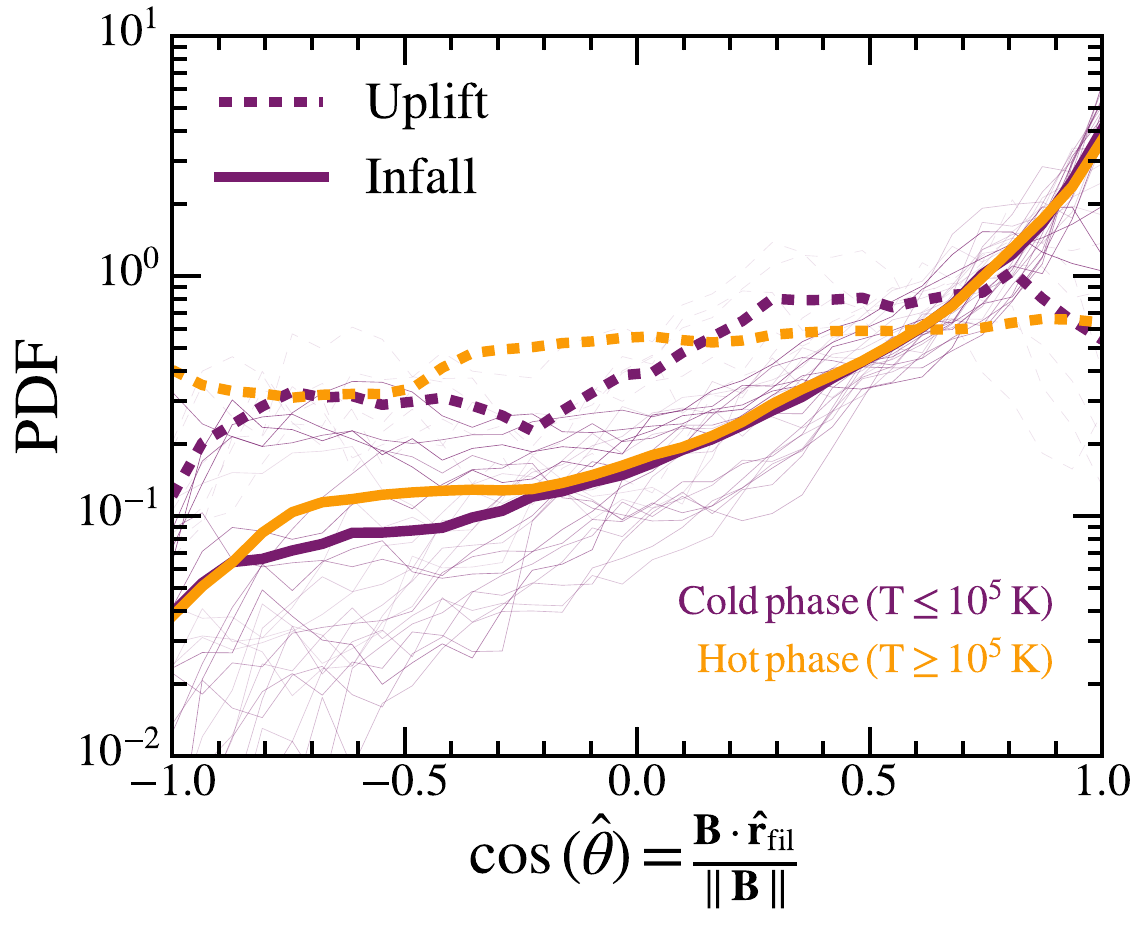}}
    \caption{Cell volume-weighted distribution of the angle between the magnetic field, $\mathbf{B}$, and the unit vector, $\mathbf{\hat{r}}_{\rm{fil}}$, for filament \#2, with $\mathbf{\hat{r}}_{\rm{fil}}$ indicating the main axis of the filament, pointing outwards from the cluster's centre. The dashed line indicates the distribution averaged over the snapshots corresponding to the phase where the average radial velocity of the filament is positive, while the solid curved is averaged on the subsequent infalling part (see Fig.~\ref{fig:FilamentsPhasePlot}). The thin curves in the background indicate the successive individual curves for the cold phase, with darker colours associated with later times.}
    \label{fig:MagAlign}
\end{figure}

\begin{figure*}[h!]
    \includegraphics[width=\textwidth]{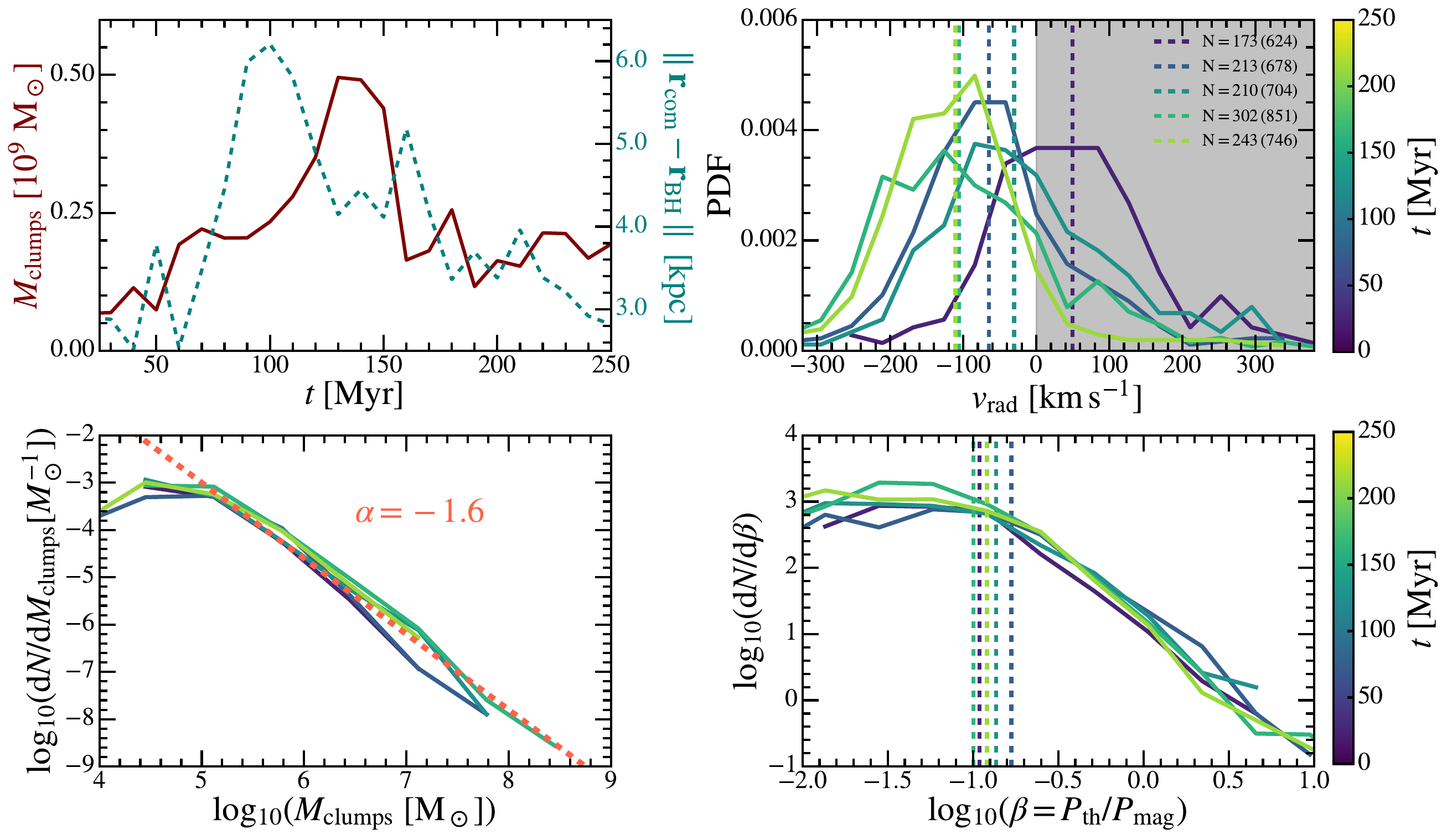}
    \caption{Evolution of the properties of the filament system (\#2) (see also Fig. \ref{fig:FilamentSequence}) at five equally spaced times. Top left panel: Evolution of the total cold mass budget of the system (solid line) and position of the centre-of-mass of the system (dashed line) as a function of time. Top right panel: Distribution of the radial velocity for various times, along with their associated median values (dashed line). The shaded grey area represents the positive radial velocity region. Lower left panel: Mass distribution of the clumps at various times. Lower right panel: Distribution of the plasma, $\beta$, at various times and associated median values (dashed line). The number, $N$, in the top right panel indicates the number of clumps found at each time, with and without (in parenthesis) the constraint on the minimum number — ten — of cells for a clump to be included in the sample.}
    \label{fig:Filament1Histo}
\end{figure*}

A sequence of images showing the formation and evolution of filament system \#2 is shown in Fig.~\ref{fig:FilamentSequence}. The panel shows, from left to right: column gas density, density-weighted radial velocity, and a mid-plane slice of the plasma, $\beta$, overlaid by magnetic field lines. At the initial state (first row), the system consists of a distribution of small, magnetically supported cloudlets and short filaments of length $\sim$ 1--2 kpc, and thicknesses close to the resolution limit. The radial velocity of this distribution of clouds shows strong internal variations depending on the location, with some clouds falling towards the centre of the cluster, and other ones exhibiting a positive radial velocity above 200 km s$^{-1}$, likely resulting from their interaction with the jet. After 40 Myr (second row), a more extended distribution of cloudlets starts gathering along the main axis of the plot, on a length scale of $\sim$ 10 kpc. The radial velocity profile varies smoothly along the distribution, from positive radial velocities at the tip of the structure to steady and infalling motions in regions closer to the centre of the SMBH. Interestingly, the clouds located at the tip of the expanding structure have velocities above 300 km s$^{-1}$ at $t=40$ Myr. Considering the gravitational field imposed in our simulation (see Sect. \ref{sect:initialconditions}), a simple integration of the equations of motion shows that the clumps should come to rest at $t\sim 40$ Myr if their motion was ballistic. Hence, the presence of some clumps with radial velocities above 300 km s$^{-1}$ after several tens of millions of years suggests that the interaction between the AGN outflows and the cold clouds lasts for several tens of millions of years, sustaining momentum transfer from the jets to the cold clouds. \\

At $t\sim$ 100 Myr, a clear filamentary structure starts to appear, which extends up to distances of $\sim 15$ kpc from the centre of the cluster. The bulk of this thread is magnetically supported, as $\beta$ drops by nearly two orders of magnitude between the ICM and the core of the filament, where its reaches values of $\beta \sim 10^{-1} - 10^{-2}$. At later times, the filament falls back towards the centre of the cluster with radial velocities down to - (200 -- 300) km s$^{-1}$. At $t=170$ Myr, the filament structure of the system vanishes. The resulting distribution of clouds looks similar to the initial state, with dozens of clumps and small-scale filaments.

The last column of Fig.~\ref{fig:FilamentSequence} suggests that the formation of the filament leads to a reshaping of the magnetic field's topology, which seems to transition between a relatively disorganised state to a topology where field lines in and around the filament are aligned with the main filament axis. To study this phenomenon more quantitatively, we present Fig.~\ref{fig:MagAlign}, which depicts the cell volume-weighted distribution of the angle between the magnetic field, $\mathbf{B}$, and the unit vector, $\mathbf{\hat{r}}_{\rm{fil}}$, for filament \#2, where $\mathbf{\hat{r}}_{\rm{fil}}$ indicates the main axis of the filament, oriented outwards from the centre of the cluster. The dashed curve indicates the distribution averaged on the snapshots corresponding to the phase where the average radial velocity of the cold gas is positive (see Fig.~\ref{fig:FilamentsPhasePlot}). The solid curves indicate the distribution in the later phase corresponding to the infall of the filament. The purple (orange) curves indicate the distribution for the cold (hot) phase, respectively. All of the considered cells are restrained here to the same region as for the analysis performed in Fig.~\ref{fig:FilamentStatistics}: a cylinder aligned with $\mathbf{\hat{r}}_{\rm{fil}}$ of radius $r_{\rm{cyl}}=2$ kpc. In the first phase, there is no clear trend visible in the distribution of the magnetic field orientation. At later times (solid curves), the distribution moves towards $\cos{(\hat{\theta})}=+1$ due to a preferential orientation of the magnetic field in both the hot and cold phases towards the main filament axis.

In Fig.~\ref{fig:Filament1Histo}, we show the evolution of several physical quantities characterising this same filament system, \#2, presented in Fig.~\ref{fig:FilamentSequence} (see also left panel of Fig.~\ref{fig:FilamentsPhasePlot}). The top left panel indicates the evolution of the cold gas mass contained in the filament (solid line), as well as the distance of its centre of mass to the centre of the cluster (dashed line). The other panels show the PDF of the clumps belonging to this filament system for various physical quantities; namely, the radial velocity of the clumps (top right panel), their mass distribution (lower left), and their plasma beta (lower) right. Time is referred here with respect to the begin of the formation of the filament; that is, when the first clumps start to move upwards (see Fig.~\ref{fig:FilamentSequence}).
The system of clumps reaches its maximum altitude at $t=100$ Myr, while the cold gas mass peaks 30 Myr later, suggesting that a large fraction of the gas (roughly 60 \%) forms through condensation when the filament falls back towards the centre of the cluster. Interestingly, both the mass and plasma beta PDF exhibit a constant shape over the course of the simulation, even though the overall morphology of the system varies widely over the life cycle of the filament. We find no clear evolution in the magnetic pressure inside the clumps. The median value of $\beta$ remains remarkably constant, around $\sim 10^{-1}$, and the proportion of the number of clumps with a dominant magnetic pressure (i.e. with $\beta < 1$) is $>$90 \% at all times.

As is shown in the top right panel, the system does not move as a rigid body, since the radial velocity PDF exhibits a large scatter, ranging from $-400$ to $+300 \, \rm{km} \, \rm{s}^{-1}$. Even when the overall bulk motion motion of the filaments falls back towards the centre of the cluster (i.e. as of $t\sim 100$ Myr), up to 30 \% of all the clumps still have a positive radial velocity, likely due to constant interactions between clumps located close to the feedback injection region and the AGN jets.

As was discussed above, the formation of filaments in our MHD run results from the combination of both upward-moving cold clumps and in situ condensation of the ICM. Since our current set-up does not include tracer particles, we cannot unambiguously disentangle the contribution of these two phenomena and we postpone such analysis to a subsequent study.

\begin{figure*}[h!]
    \includegraphics[width=\textwidth]{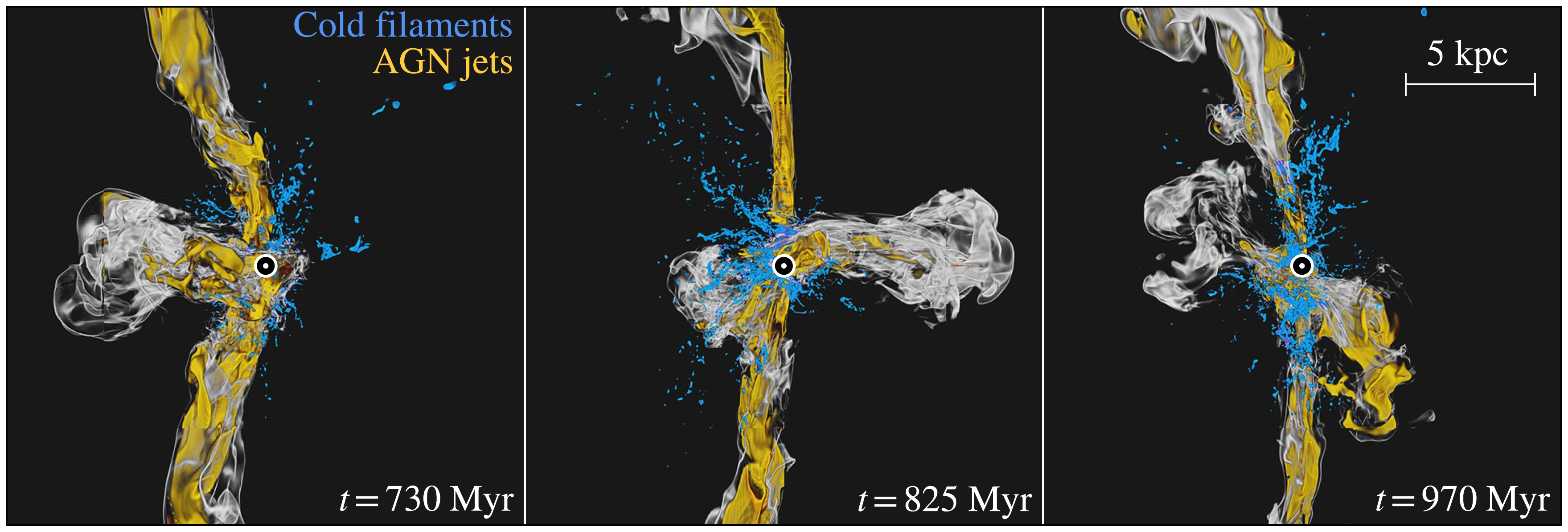}
    \caption{Volume-rendering of the inner (20 kpc)$^3$ of our fiducial MHD run at three different times. The yellow and white components represent the isodensity surfaces for $3 \times 10^{-3} \ \rm{cm}^{-3}$ and $2 \times 10^{-2} \ \rm{cm}^{-3}$, respectively, and trace the outflowing AGN jet material. The blue component traces the cold gas of density $20 \ \rm{cm}^{-3}$. The figure shows that, even though the AGN jets are accelerated along the vertical axis of this image, interaction with the dense infalling filaments leads to the formation of cavities that move in directions strongly inclined relative to the main jet axis. The round black and white marker at the centre of each panel represents the position of the SMBH particle. (\href{https://www.youtube.com/playlist?list=PLe_DeluPkGwM2tXKomlp8vUiQOXuYhKre}{Movies available})}
    \label{fig:volumetric}
\end{figure*}

\section{Impact of cold gas on the active galactic nucleus outflows}
\label{Impact}

Cavities in galaxy clusters have been widely studied using X-ray observations. Typically, pairs of cavities with radii of $\sim 10$ kpc are found up to a few tens of kiloparsecs away from the central galaxy. These are sometimes filled with radio emission associated with AGN radio jets that inflate the cavities \citep{Boehringer_1993,Birzan_2004}. Understanding the processes shaping these cavities is critical to assessing how AGN feedback can balance the cooling of the ICM. In previous works, it has been discussed that high-resolution simulations (typically with maximally refined cells smaller than $\sim$ 100 pc) tend to form morphologies closer to FR I galaxies, with strongly collimated jets that propagate out to distances of up to hundreds of kiloparsecs. A direct consequence is a less isotropic heating of the ICM and possibly a failure of the AGN feedback to prevent the catastrophic cooling of the ICM (as was discussed in \citet{Li_2014} and \citet{Verlaneo_2006}). In previous works, this problem has been addressed by including jet re-orientation, either by enforcing jet precession around a fixed axis \citep{Li_2014,Meece2017} or by computing the jet axis following a spin-driven approach \citep{Beckmann_2019, Beckmann_2022}. In \citet{Ehlert_2023}, the cold material is found to deflect the gas, inflating large bubbles that are eventually no longer aligned with the injection axis. In this section, we explore the effects of the small-scale jet-filament interaction on the overall morphology of the outflowing jets and discuss the morphology of the outflows. \\

\subsection{Deflection of the jets by cold gas}
\label{interaction}

We find numerous occurrences of interactions between the infalling cold gas and the outflowing AGN jets. Since the density contrast between the injected hot material and the clouds is up to $\sim$ 6 orders of magnitude, the light jets are easily deflected by the filaments.

In Fig.~\ref{fig:volumetric}, we show a volume-rendering of the inner (20 kpc)$^3$ of our cluster. The yellow and white components represent the isodensity surfaces for $3 \times 10^{-3} \, \rm{cm}^{-3}$ and $2 \times 10^{-2} \, \rm{cm}^{-3}$ and trace the outflowing AGN jet material. The blue component traces the cold gas of density $20 \, \rm{cm}^{-3}$. Interactions of the cold gas with the hot material accelerated by the AGN jet lead to the formation and propagation of cavities in all directions relative to the vertical jet axis. In some cases, the distribution of the infalling cold gas splits both jets simultaneously, leading to the formation of a second pair of cavities (right panel). \\

\begin{figure*}[h!]
    \includegraphics[width=\textwidth]{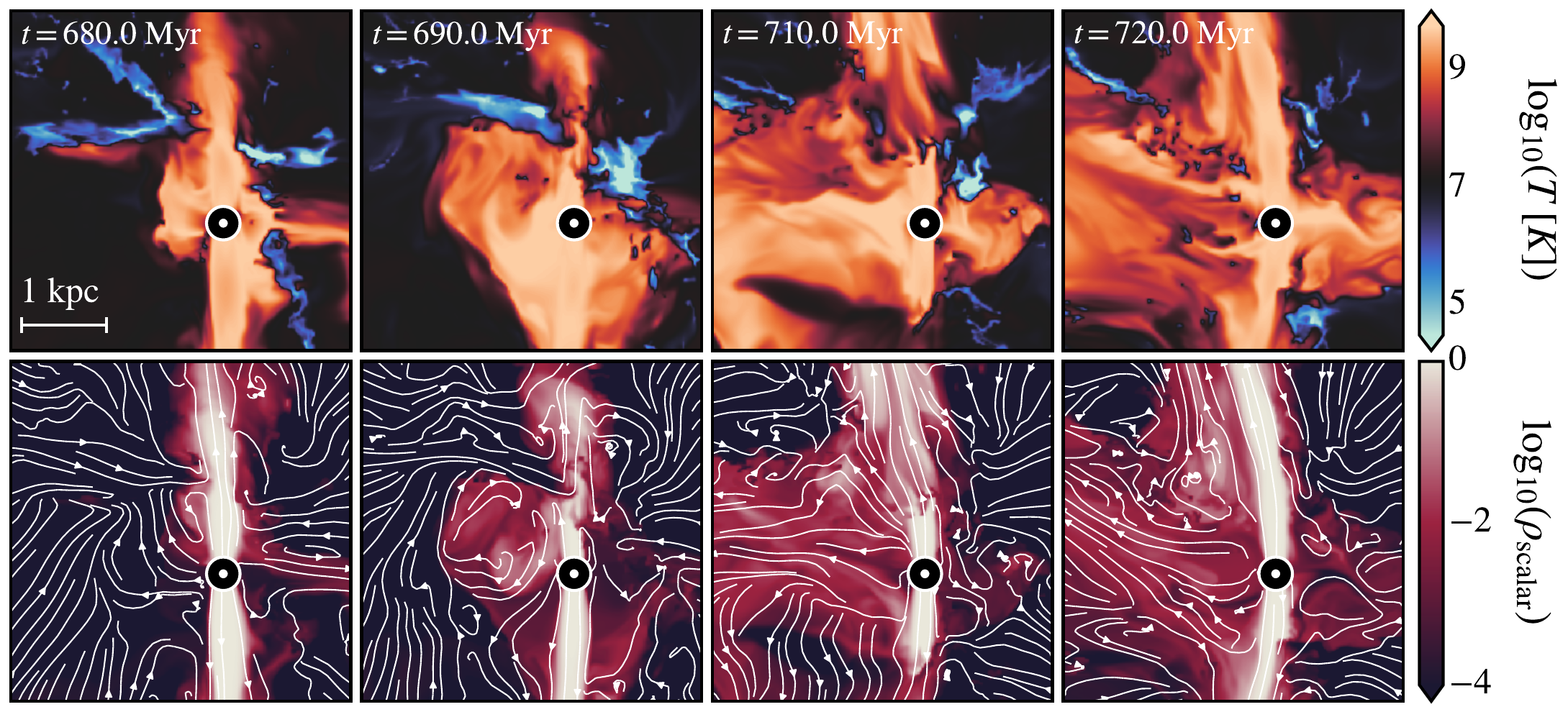}
    \caption{Density-weighted projections of the temperature field (top panels) and passive jet scalar (bottom panels), overlaid with the velocity streamlines of the gas. The depth of projection is 250 pc. The circular marker indicates the position of the SMBH particle and the injection axis of the jet is the vertical axis of the image. The interaction of infalling filaments with the AGN outflows leads to a deflection of the upper jet and the formation of a cavity propagating in a direction orthogonal to the main jet axis. (\href{https://www.youtube.com/playlist?list=PLe_DeluPkGwM2tXKomlp8vUiQOXuYhKre}{Movies available})}
    \label{fig:deflection}
\end{figure*}

In Fig.~\ref{fig:deflection}, we show a projection of the density-weighted temperature (top panels) and passive jet scalar (bottom panels). The image sequence in Fig.~\ref{fig:volumetric} shows the formation of the cavity visible in the left panel. The projection depth of 250 pc is set to the typical jet diameter in the inner few kiloparsecs of the box, allowing us to unambiguously visualise the interaction between the cold gas and the jets. The velocity stream lines are overlaid on the temperature maps and the round marker indicates the position of the black hole. As is shown in the temperature maps, several cold filaments converge towards the upper jet (visible in blue in the first column). The top left filament starts to interact with the jet in the second column and leads to a splitting of the jet. Part of the hot material is strongly deflected and starts carving a cavity orthogonal to the main jet axis. The filament is shattered from its interaction with the jet and breaks into smaller clumps, visible in the last two columns. By $t=720$ Myr, all the remaining clumps in the way of the jet have been destroyed. \\

\subsection{Morphology of the outflows}

\begin{figure}[h!]
    \centering
    \resizebox{\hsize}{!}
    {\includegraphics{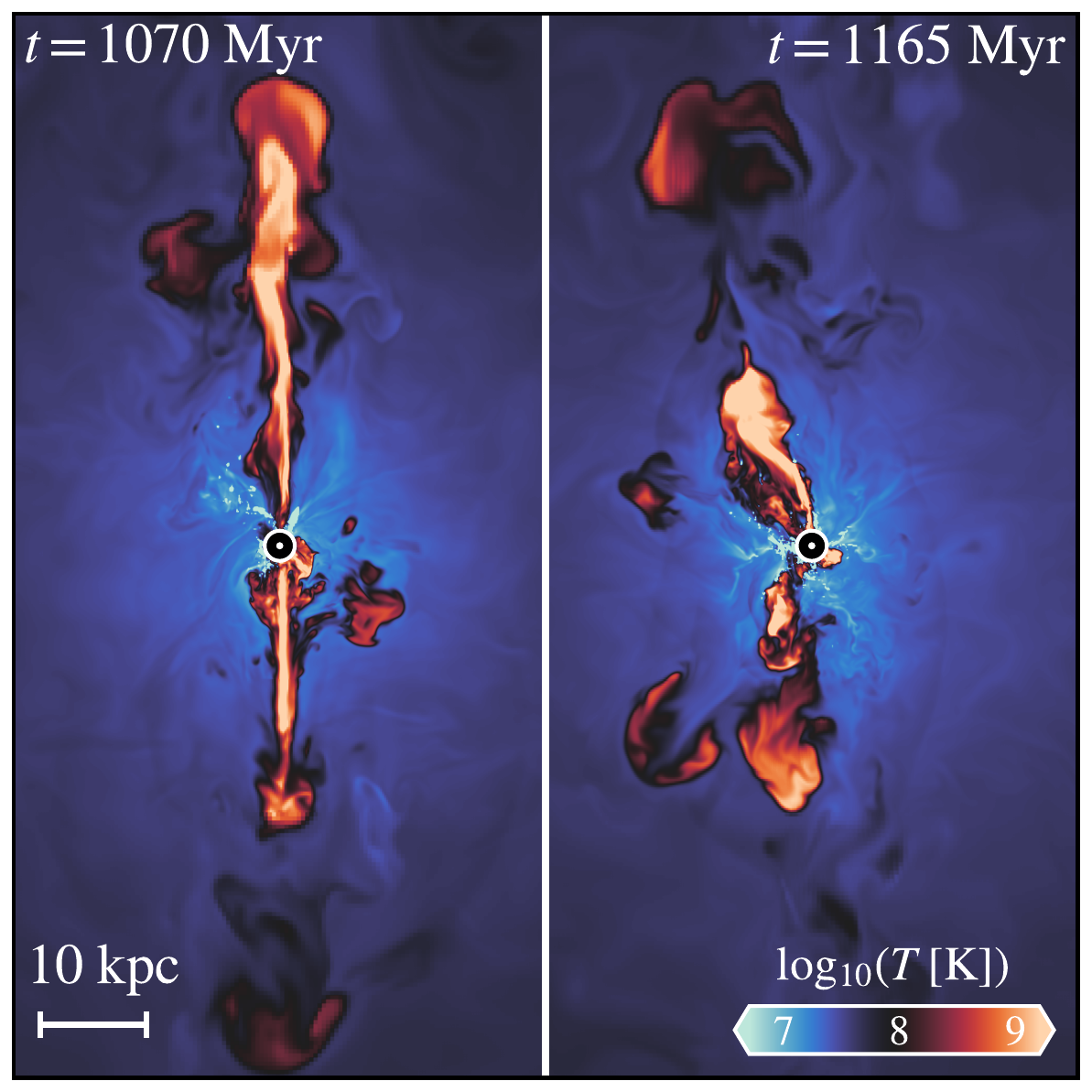}}
    \caption{Projection of the temperature field in the mid-plane of the simulation at two different times. The left panel shows a typical example of jets propagating in an unperturbed way, staying collimated up to a distance of 45 kpc. In the right panel, several hot spots resulting from previous AGN bursts can be identified in various directions, as can jet bending. The circular marker represents the position of the SMBH particle, and the jet injection axis is the vertical axis of the image.}
    \label{fig:collimation}
\end{figure}

Cold gas is not always found to lie along the jet axis where it could perturb the jets. Typical examples for each of these two behaviours are presented in Fig.~\ref{fig:collimation}, showing a density-weighted projection of the temperature field in a narrow, 500 pc-deep region centred on the mid-plane of the simulation. In the left panel, the jets exhibit a strongly collimated morphology at distances of up to $\sim 45-50$ kpc. The right panel shows the same region of the simulation nearly 100 Myr later. Both jets are bent due to their interaction with the clumps and several hot spots that formed during previous AGN bursts are visible. These are propagating in directions different from the jet axis. Quantifying the relative duration of these two regimes and their possible impact on the ICM heating is beyond the scope of this paper.\\

\begin{figure*}[h!]
    \includegraphics[width=\textwidth]{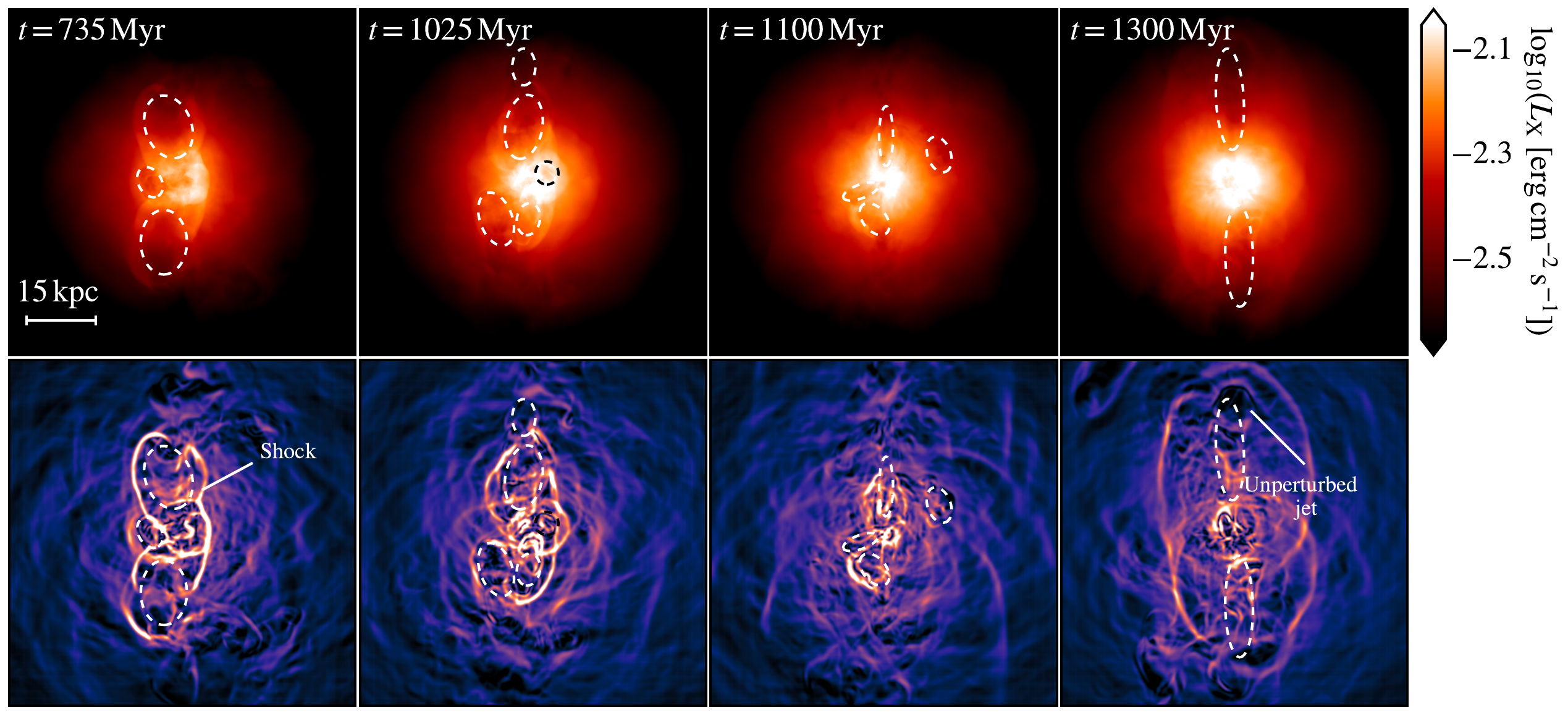}
    \caption{X-ray mock images of our cluster at the native resolution, i.e. not matched to a particular X-ray instrument. The top row shows the X-ray luminosity in the 2--7 keV range. The dashed lines indicate the position of the cavities. The lower row is comparable to Fig. 2 from \citet{Sanders_2016} and shows the corresponding GGM, revealing shocks, ripples, and cavities.}
    \label{fig:Xray}
\end{figure*}

Cavities seen in X-ray observations of galaxy clusters show diverse morphologies and spatial scales. The typical diameter and cluster-centric distance of cavities can range from a few kiloparsecs to extreme values of hundreds of kiloparsecs, such as in MS 0735.6+7421 \citep{Larrondo_2022}. In some cases, more than one pair of bubbles is observed. These are probably associated with successive AGN bursts, such as in the Perseus cluster \citep{Fabian_2006}, NGC 5813 \citep{Larrondo_2022}, or RBS 797 \citep{Ubertosi_2023}. In order to compare the appearance of our simulation in X-rays to observations, we produced mock images with the {\scshape{yt}} visualisation package. In Fig.~\ref{fig:Xray}, we show a series of X-ray mock images at four different times in our high-resolution MHD simulation. The top row shows the X-ray luminosity in the 2--7 keV range for the inner 50 kpc of the box, while the bottom row shows the corresponding Gaussian gradient magnitude (GGM) in the inner 100 kpc of the box. The GGM image emphasises faint luminosity variations in the X-ray map and reveals shocks, cavities, and ripples (see Fig. 2 from \citet{Sanders_2016} for comparison). The specific times were selected to showcase the diversity of X-ray structures found in our simulation.

The first column shows an example of a pair of cavities. Their typical height and width is roughly 10 kpc and their aspect ratio is close to unity. Such a morphology is qualitatively similar to the pair of cavities found in the innermost region of the Perseus cluster \citep{Fabian_2006}. There is an additional, smaller cavity resulting from the deflection of the jet by infalling filaments close to the AGN, with a typical diameter of 5 kpc. Its position and relative size can be compared to the second pair of cavities observed in RBS 797 (see Fig.~7 from \citet{Ubertosi_2023}), although it is unclear whether these have a similar physical origin. It should be noted that the formation of this small cavity has been presented in Sect.~\ref{interaction} and Fig.~\ref{fig:deflection}. The shock wave enveloping the bubbles is also visible in the GGM figure. The third and fourth columns show typical occurrences of jet propagation without jet deflection. In that case, the jets propagate along a straight path and carve narrow channels of under-dense gas into the ICM. 

\section{Discussion}
\label{Discussion}

We have achieved spatial resolutions close to a cloud-crushing set-up in a simulation of an idealised cluster with self-regulated AGN feedback. We have shown that the inclusion of magnetic fields with a high resolution has a significant impact on the morphology and dynamics of cold gas in a jet-regulated ICM. Our study, however, remains imperfect and several questions remain open. \\

\subsection{Upward-moving gas and uplifting}

Various channels for the formation of the filaments have been discussed in the literature. \citet{Yuan_2014b} and \citet{Voit_2017} discuss a picture in which low-entropy hot gas is uplifted through interaction with the AGN jets. This uplifted hot gas then condenses and builds up mass, while decelerating and falling back towards the centre of the accretion zone. In \citet{Prasad_2015}, the jets are found to effectively uplift clumps of cold gas, which might rise up to altitudes of a few tens of kiloparsecs. Since our simulations do not include tracer particles, we cannot unambiguously quantify the relative importance of these two formation channels and our interpretation can only rely on simple considerations. Throughout the whole duration of our fiducial MHD run, the inner 3--4 kpc of the simulated cluster constantly contains a myriad of clumps and short filaments (with a typical length of a few hundred parsecs) of cold ($T < 10^{5}$ K) gas. The first few kiloparsecs of the path of the jets thus consist of a strongly multi-phase medium, and interactions between the jets and cold clumps of gas are frequent, as is discussed in Sect.~\ref{Impact}. Looking at the kinematics of cold gas in the direction of propagation of the jets, we found ubiquitous cold gas structures with positive radial velocities of up $\sim 10^3 \, \rm{km}\,\rm{s}^{-1}$ at distances as small as a few hundred parsecs away from the kinetic jet injection region. This is an indication that the presence of fast outflows of cold gas is likely the result of a physical uplifting by the jets. While there remains a possibility that the origin of these fast upward-moving clumps could be explained by the condensation of low-entropy hot gas, this condensation then must happen on distances as small as a few hundred parsecs, roughly an order of magnitude below results from previous work \citep{Yuan_2014b}.

\subsection{Length of the filaments}

We note that our filaments have relatively short lengths compared to the extents of optical nebulae in some CC clusters. In the Perseus cluster, H$\alpha$-emitting filaments are observed at distances of up to 50 kpc from the centre of the BCG \citep{Salome_2006}. The typical extents of the cold features observed in our simulation range between $\sim$5 and $\sim$15 kpc, closer to the H$\alpha$ filaments observed in the Centaurus cluster \citep{Olivares_2019}. One possible explanation is that the use of static grid refinement leads to an under-estimate of cooling at larger radii, where the cells are coarser. However, we note that other simulations using adaptive mesh refinement also find similar filament lengths \citep{Beckmann_2019}. It is also likely that modelling the relativistic nature of the jets and resolving gas with increased spatial resolution near the core of the cluster will result in a more effective coupling between the jets and the ambient gas, leading to higher launching velocities and higher altitudes being reached from the centre of the cluster, potentially triggering condensation at larger radii. \\

\subsection{Magnetic field strength in the filaments}

It has been a matter of discussion whether the kiloparsec-scale filaments of cold gas observed in the ICM of clusters and in the circumgalactic medium of some elliptical galaxies are magnetically supported. In \citet{Fabian_2008}, a minimum magnetic field strength of 25 $\mu$G is found to be required to stabilise the filaments observed in NGC 1275 against tidal shear. Similar field strengths are also obtained from Faraday rotation measurements of the magnetic field strength in the inner kiloparsecs of the Perseus cluster \citep{Taylor_2006}. Measurements of the magnetic field strength of the other systems, such as the cold gas filaments in the Virgo cluster, lead to inferred values of up to $\sim 100 \ \mu$G \citep{Werner_2013}.
In our MHD run, we find that the magnetic field strength of the clumps and filaments is particularly high. While the typical strength values for our filaments are around 80--100 $\mu$G, we also find individual cells or small clumps with values up to 200--250 $\mu$G. While similar maximum values have been found in MHD cloud-crushing simulations \citep{Jennings_2023} as well as in self-regulated cluster simulations \citep{Ehlert_2023}, it remains unclear whether such values could be reached in real CC clusters, since observations favour values of up to $\sim 50-100 \, \mu$G. It is possible, though, that external sources of pressure such as cosmic rays may reduce the overall magnetic field strength in the threads.

\subsection{Jet propagation and off-axis cavities}

As is discussed in Sect. \ref{Impact}, our high-resolution MHD run produces many cavities that are misaligned with the jets injection axis. Although a similar behaviour has been observed and/or discussed in previous works \citep{Li_2014,Qiu_2019,Wang_2020,Wang_2021,Ehlert_2023}, we find that these deflected cavities are relatively short-lived and expand on much shorter distances than the outflows propagating in directions aligned with the main jet axis, typically up to distances of $\sim 10$ kpc. A possible explanation for this behaviour is our spatial resolution, which allows us to resolve the fragmentation of the cold filaments into smaller clumps. The deflection of the jet is then only partial, with a significant fraction of the material still propagating along the injection axis. We find our low-resolution runs to exhibit behaviour closer to that observed in \citet{Ehlert_2023}, with larger cavities fully deflected in nearly any direction. We emphasise that we assume a sub-relativistic propagation of our jets, a hypothesis that might not hold for typical length scales of $\sim 1$ kpc. Previous studies have shown that relativistic jets with powers above $\sim 10^{44} \ \rm{erg} \, \rm{s}^{-1}$ are less affected by the presence of a clumpy interstellar medium in disc galaxies \citep{Tanner_2022}, which might reduce the deflection by the cold gas even further. However, denser molecular gas that would form if we allowed gas to cool below our temperature floor at $10^{4.2}$ K could lead to more deflection. We postpone the study of relativistic effects and the impact of modelling colder gas to further investigations. Cavities such as those presented in the first two columns of Fig.~\ref{fig:Xray} are common in our simulations, and are consistent with the typical sizes of observed X-ray cavities in galaxy clusters \citep{Panagoulia_2014}. Whether the collimated jet morphology presented in the last two columns has a correspondence in observations is less clear. This is perhaps an artefact of our set-up, however, since ambient bulk motions in real galaxy clusters may prevent such elongated cavities. It could also indicate additional magnetic support, a wider jet opening, and/or jet re-orientation.

\subsection{Additional physics}


Anisotropic thermal conduction can change the dynamics and morphology of the multi-phase gas. Thermal conduction is mediated by electrons from the hot surrounding medium that penetrate into the cold filaments. This has been studied analytically \citep[e.g.]{Cowie_1977} and with simulations \citep{Jennings_2023,2023ApJ...951..113B}. In the case of anisotropic conduction, the impact of conduction has been shown to be limited and depends on the field orientation and degree of tangling.


Another physical process that has received attention in recent years is heating by cosmic rays, which can play an important role in clusters. Radio observations show that cosmic rays pervade at least the inner regions of galaxy clusters. For instance, \cite{2017ApJ...844...13R} found that cosmic-ray streaming was essential for AGNs to regulate the cooling of the ICM (see also \cite{2023A&ARv..31....4R} for a recent review on the theory of cosmic ray feedback in galaxies and clusters of galaxies). While the modelling of cosmic ray feedback is complicated by the largely unknown parameters that govern cosmic-ray transport, there are clearly avenues for future research in this area.


The overall morphology of the cold gas could also be impacted by a re-orientation of the jet injection axis. \citet{Beckmann_2019} follow a spin-driven approach in a purely hydrodynamic set-up and find large and fast variations in the jet orientations, up to $\dot{\theta}_{\rm{jet}} \sim 10^{\circ} \, \rm{Myr}^{-1}$. As a consequence, the ICM is propelled more isotropically by the jets and the feedback power exhibits strong variations with peaks up to $10^{49} \, \rm{erg} \, \rm{s}^{-1}$, orders of magnitude above the maximum AGN power found in our runs. Whether such extreme re-orientations of the black hole spin axis are achievable in reality is a matter of debate \citep{Nixon_2013}.


It has been shown that threads of molecular gas spatially correlate with the location of the H$\alpha$-emitting nebula observed near the BCG of the Perseus cluster, and it is believed that gas in those regions could reach temperatures as low as 10--100 K \citep{Salome_2006}. In our current set-up, we impose a temperature floor of $10^4$ K, which is likely to limit the total amount of cold gas formed and its maximum density \citep{McDonald_2010}. Below such a temperature threshold, it is suspected that photoionisation from young stars could play a role in the powering of the filaments, and should thus be included in the cooling calculation, either with explicit radiative transfer or additional heating source terms in the radiative cooling algorithm.

\section{Conclusions}
\label{Conclusion}

The main conclusions of our work are as follows:

\begin{itemize}

    \item We confirm that magnetic fields have an important impact on the dynamics of the cold gas. The structures obtained in our MHD simulation resemble the morphology of the H$\alpha$ filaments observed in CC clusters on kiloparsec scales, such as the Centaurus or Perseus clusters. In the absence of magnetic fields, simulations predominantly produce clumps instead of filaments with a lower specific angular momentum.
    
    \item The jet lifts up gas and promotes in situ cooling. This leads to cold gas formation at the edges of the hot, outflowing bubbles or jets. As much as a few tens of percent of the mass of some filaments is carried by upward-moving cold clumps with velocities of up to $\sim10^3 \,~\rm{km}\,~\rm{s}^{-1}$, while the remainder condenses out of the hot phase during the fall back onto the AGN.
     
    \item Magnetic pressure largely dominates thermal pressure in the inner regions of the filaments. The values for plasma-$\beta$ go as low as $10^{-1}-10^{-2}$, and the magnetic field strengths in the filament core are typically around 100 $\mu$G.
     
    \item Cold filaments in our high-resolution MHD run are not monolithic. Their internal structure is complex and contains hundreds of sub-structures such as clumps with thicknesses down to a few tens of parsecs. Using a clump-finding algorithm, we isolated several distinct filaments and studied their internal compositions individually. While filaments vary in length and mass, the overall distribution of their sub-structures is constant across filaments. In particular, the clumps population within each filament is found to follow a $\mathrm{d}N/\mathrm{d}M \propto M^{-1.6}$ mass distribution, in agreement with other simulation-based studies.
    
    \item We have identified the cyclic life of some filaments. We find that small clumps are initially moving upwards, likely through jet interaction. The filament structure becomes visible as the clumps reach their highest altitude. The filamentary structures are preferably found to be infalling towards the centre of the cluster.
    
    \item The formation of filaments lead to a re-organisation of the magnetic field topology in the hot and cold phases. While the topology of the magnetic field has no preferential orientation prior to the filaments' formation, field lines gradually align with the main filament axis during the formation sequence and point outwards from the cluster's centre.
    
    \item Jet-cloud interactions contribute to the formation of large cavities. Infalling threads and clumps of cold gas lying along the main jet axis efficiently deflect and scatter the hot gas, favouring the inflation of bubbles. While jet-cloud interactions are frequent, we also find transient AGN bursts with no deflection. In that case, the jet remains collimated out to distances of tens of kiloparsecs and its morphology resembles that of an FR I galaxy.
    
    \item Off-axis cavities are ubiquitous in our MHD runs. The large density contrast between the cold gas and the jet material leads to the formation of cavities propagating in any direction, including orthogonal to the jet axis. These cavities are relatively small, with a typical diameter of a few kiloparsecs. This is likely due to the fact that most of the filaments are too narrow to deflect the jet entirely, and thus only a fraction of the injected material participates in this process.  In purely hydrodynamic runs, cavities are oriented along the jet axis.

\end{itemize}

\begin{acknowledgements}
MF thanks Yu Qiu, Ricarda Beckmann, Marine Prunier, Deovrat Prasad and Ben Wibking for insightful discussions, as well as Jacob Shen and Evan P. O'Connor for providing help with visualisation packages. \\
This project has received funding from the European Union’s Horizon 2020 research and innovation programme under the Marie Skłodowska-Curie grant agreement No 101030214.\\
MB acknowledges funding by the Deutsche Forschungsgemeinschaft (DFG, German Research Foundation) under Germany's Excellence Strategy -- EXC 2121 ``Quantum Universe'' --  390833306 and project number 443220636 (DFG research unit FOR 5195: "Relativistic Jets in Active Galaxies").\\
BWO acknowledges support from NSF grants \#1908109 and \#2106575, NASA ATP grants NNX15AP39G and 80NSSC18K1105, and NASA TCAN grant 80NSSC21K1053. \\
The authors gratefully acknowledge the Gauss Centre for Supercomputing e.V. (www.gauss-centre.eu) for funding this project by providing computing time through the John von Neumann Institute for Computing (NIC) on the GCS Supercomputer JUWELS at J\"ulich Supercomputing Centre (JSC). \\
This research used resources of the Oak Ridge Leadership Computing Facility at the Oak Ridge National Laboratory, which is supported by the Office of Science of the U.S. Department of Energy under Contract No. DE-AC05-00OR22725. These resources were provided by as part of the DOE INCITE Leadership Computing Program under allocation AST-146 (PI: Brian O'Shea).\\
All simulations were performed using the public MHD code {\href{https://github.com/parthenon-hpc-lab/athenapk}{\scshape{AthenaPK}}}, which makes use of the \href{https://github.com/kokkos/kokkos}{{\scshape{Kokkos}}} \citep{Kokkos} library and the \href{https://github.com/parthenon-hpc-lab/parthenon}{{\scshape{Parthenon}}} adaptive mesh refinement framework \citep{Parthenon}. All data analysis was performed with \href{https://yt-project.org/}{{\scshape{yt}}} \citep{Turk_2011,yt4}, \href{https://matplotlib.org/}{{\scshape{Matplotlib}}} \citep{Matplotlib}, \href{https://numpy.org/}{{\scshape{Numpy}}} \citep{Numpy}, \href{https://seaborn.pydata.org/}{{\scshape{Seaborn}}} \citep{Seaborn} and \href{https://github.com/JBorrow/swiftascmaps}{{\texttt{swiftascmaps}}}. We thank their authors for making these software and packages publicly available.

\end{acknowledgements}

\bibliographystyle{aa} 
\bibliography{biblio} 

\clearpage

\begin{appendix}

\section{The impact of spatial resolution}
\label{spatialresolutionstudy}

To study the effect of spatial resolution on our results, we run additional simulations with fewer refinement levels whose parameters are summarised in Table \ref{table:parametersappendix}. The extent of each remaining refinement level, as well as the initial conditions are unchanged from one run to an other. The size of the injection region for the AGN feedback is adapted to the resolution of the cells at the centre of the box, and thus vary from one run to an other (see Sect. \ref{agnfeedback}).

\begin{table}[h!]
\caption{Extended version of Table \ref{table:parameters} including lower resolution test simulations.}
\label{table:parametersappendix}
\centering
\begin{tabular}{lccccc}\hline\hline
Name        & \begin{tabular}[c]{@{}c@{}}Max. res.\\ {[}pc{]}\end{tabular} & $f_K$ & $f_T$ & $f_M$ & \begin{tabular}[c]{@{}c@{}}$\langle B(r=0) \rangle$\\ {[}$\mu$G{]}\end{tabular} \\ \hline
hydro\_xlow & 200                                                          & 0.75  & 0.25  & -     & -                                                                               \\
hydro\_low  & 100                                                          & 0.75  & 0.25  & -     & -                                                                               \\
hydro       & 25                                                           & 0.75  & 0.25  & -     & -                                                                               \\ \hline
mhd\_xlow   & 200                                                          & 0.74  & 0.25  & 0.01  & 10                                                                              \\
mhd\_low    & 100                                                          & 0.74  & 0.25  & 0.01  & 10                                                                              \\
mhd\_mid    & 50                                                           & 0.74  & 0.25  & 0.01  & 10                                                                              \\
mhd         & 25                                                           & 0.74  & 0.25  & 0.01  & 10 \\ \hline
\end{tabular}
\end{table}

\begin{figure}[h!]
    \centering
    \resizebox{\hsize}{!}
    {\includegraphics{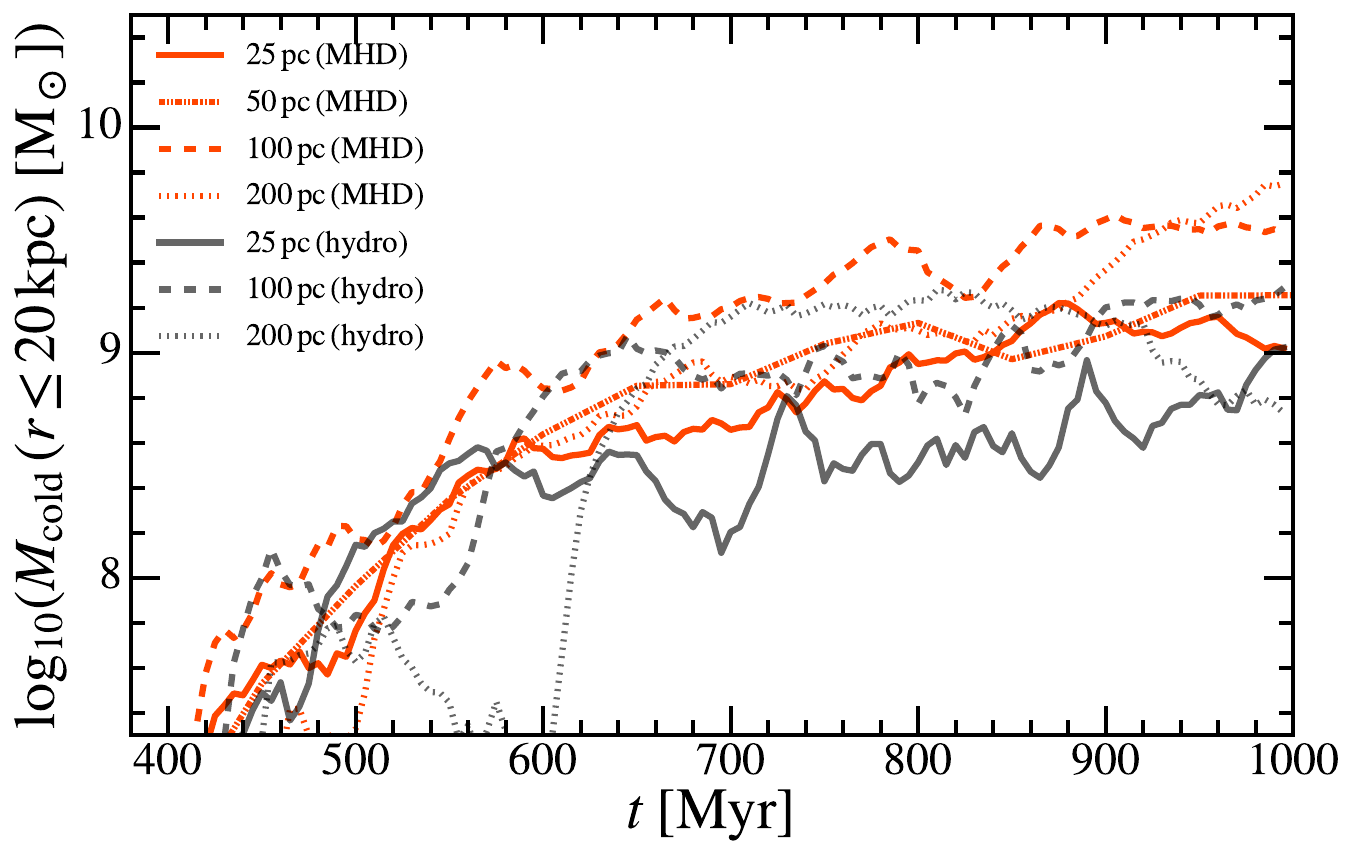}}
    \caption{Total amount of cold gas contained in the inner 40 kpc of the simulated box.}
    \label{fig:ColdGasRes}
\end{figure}

Most MHD simulations of AGN feedback have achieved spatial resolutions of no more than a few hundred pc \citep{Beckmann_2022,Wang_2021,Ehlert_2023}. We find that increasing spatial resolution to less than 100 pc affects the morphology of the cold gas, its extent, the subsequent interactions with the AGN jets, and the morphology of the outflows. In Fig.~\ref{fig:ColdGasRes}, we present the total cold gas mass in the inner 40 kpc of our box as a function of time for our 7 runs. Considering the analysis performed in this paper mostly relies on our highest resolution MHD simulation, we mostly focus on discussing resolution dependence of our MHD runs. For comparison, we also show the results for purely hydrodynamic runs.

As visible, lower resolution runs produce on average more cold gas than higher resolution ones, and can vary by up to roughly one order of magnitude (e.g. comparing 25 pc and 200 pc MHD runs at $t=1.0$ Gyr). We also note that our 50 pc and 25 pc MHD runs are producing comparable quantities of cold gas with variations less than a factor of two throughout the whole duration of the test run, and converges to $\sim10^{9} \, \rm{M}_{\odot}$ around $t\sim1$ Gyr.

\begin{figure}[h!]
    \centering
    \resizebox{\hsize}{!}
    {\includegraphics{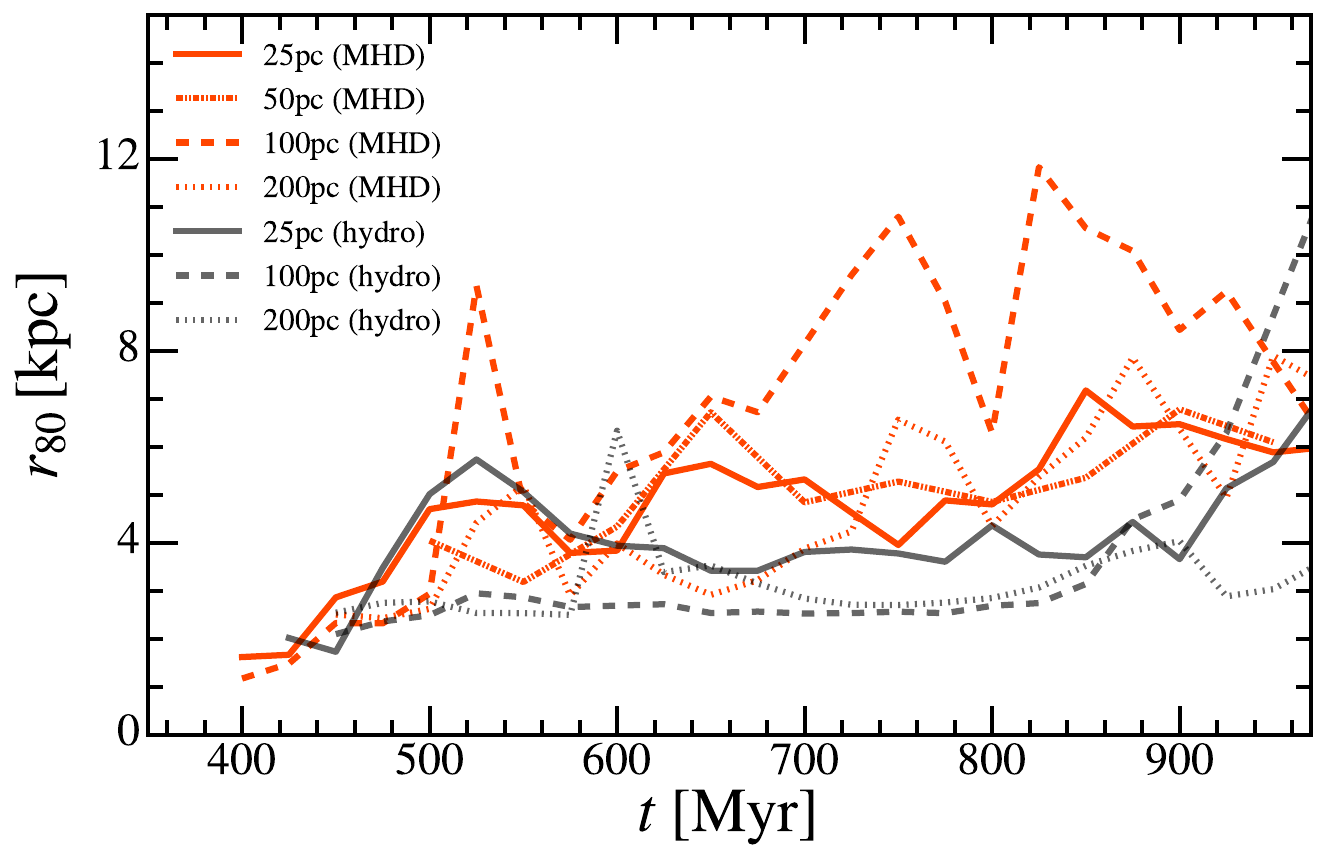}}
    \caption{Evolution of $r_{80}$, the radius containing 80\% of the cold gas in our various runs.}
    \label{fig:r80}
\end{figure}

In Fig.~\ref{fig:r80}, we show the evolution of $r_{80}$, the radius containing 80\% of the cold gas to quantify its spatial extent. Our 50 pc and 25 pc resolution MHD runs are broadly consistent, with $r_{80}$ converging to $\sim 5$ kpc, in agreement with previous work (see notably Fig. 8 from \citet{Beckmann_2022}). Interestingly, our 100 pc resolution MHD run seems to lead to a more extended distribution of cold gas, which we find to be also preferably located along the jet injection axis.

Identifying the exact physical or numerical processes leading to these spatial resolution dependencies is beyond the scope of this study as numerous possible explanations could be investigated. In particular, jet properties are not identical from one resolution run. As the spatial resolution increases, the size of the injection region is decreased to fulfill $r_{\rm{jet}}=4\,\Delta x_{\rm{min}}$ and $h_{\rm{jet}}=2\,\Delta x_{\rm{min}}$. As this might result in jet injection velocities above the velocity ceiling value $v_{\rm{ceiling}}=0.05\,c$, we enforce energy conservation by increasing the density of the injected jet material. This leads to jets in our 25 pc runs to be heavier than in our 200 pc runs by a factor of up to 2--3. This is likely affecting their propagation properties and the way turbulence is deposited in the ICM. We however emphasise that the jets remain relatively light in every run, with densities in the injection region always of the order of $\sim 10^{-26} \, \rm{g}\,\rm{cm^{-3}}$ ($\sim 10^{-2} \, \rm{cm^{-3}}$) at most. \\

\begin{figure*}[h!]
    \centering
    \includegraphics[width=\textwidth]{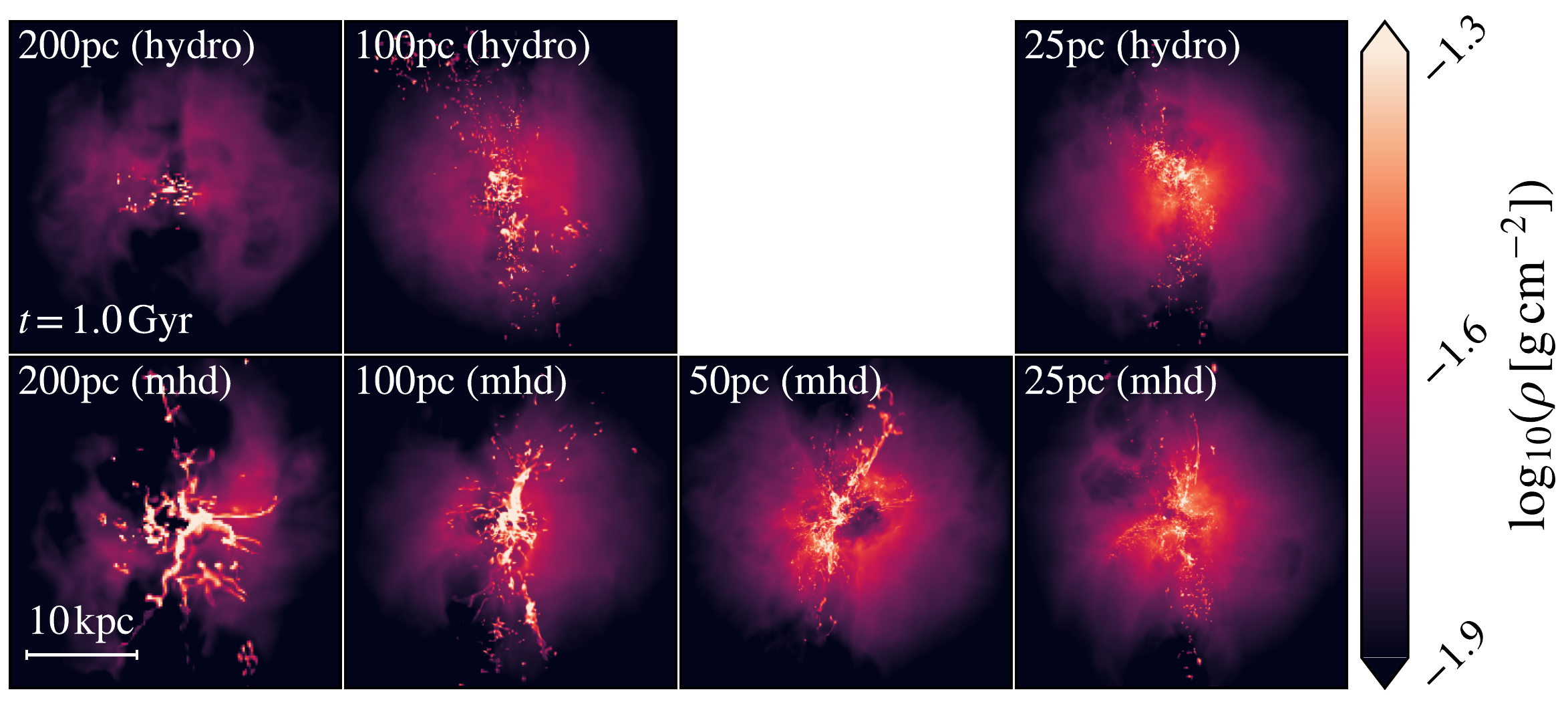}
    \caption{Gas density projection in the inner 30 kpc of our simulated cluster for each of our simulations.}
    \label{fig:ColdGasDistribRes}
\end{figure*}

In Fig.~\ref{fig:ColdGasDistribRes}, we show projections of the gas density in the inner 30 kpc of our simulated box for our various runs. Numerous filaments are visible in each of our MHD runs, although their sub-structures are unresolved for resolutions coarser than 100 pc. No clearly filamentary structures such as the ones presented in Sect. \ref{section:morphology} are found throughout the duration of our purely hydrodynamic simulations apart from a few short-lived and relatively small filaments (i.e. with typical lengths shorter than 5 kpc, thicknesses close to resolution limit) visible in our low resolutions hydrodynamic runs. \\

\section{The impact of the clump-finding algorithm}
\label{impactclumpfinder}

\begin{figure}[h!]
    \centering
    \resizebox{\hsize}{!}{\includegraphics{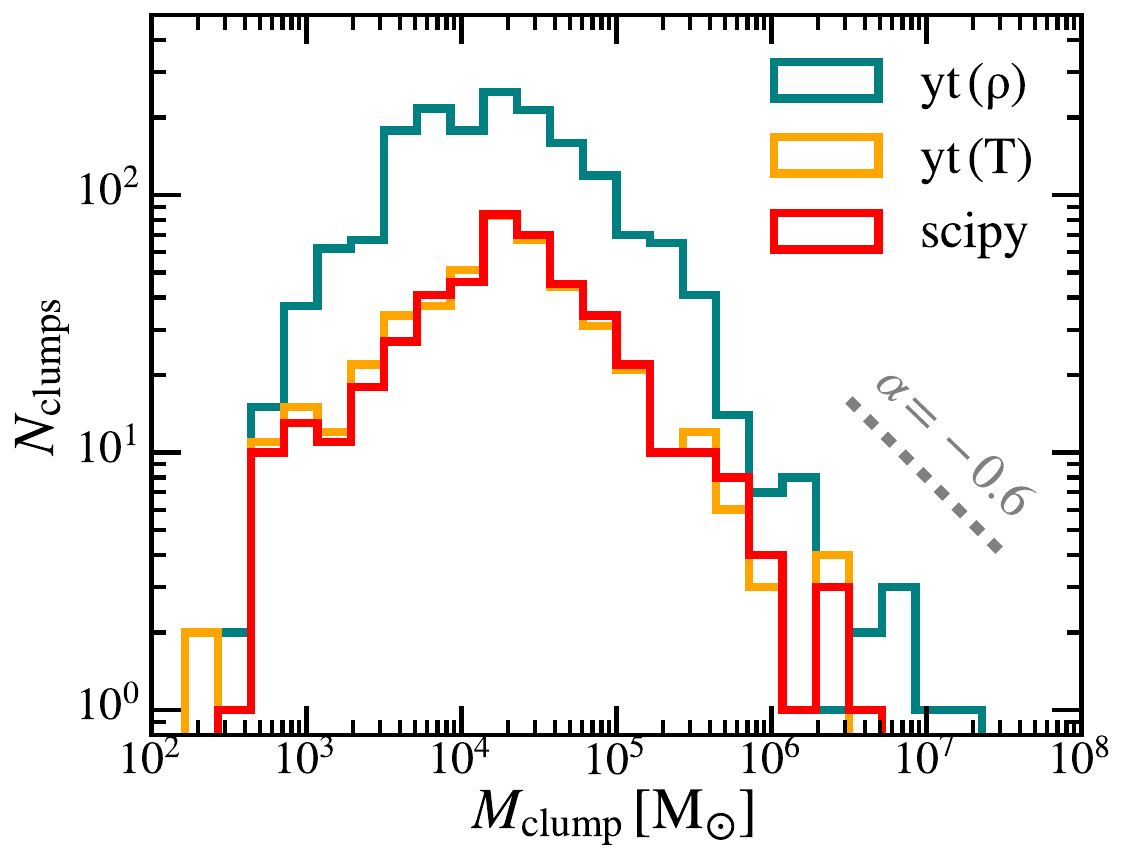}}
    \caption{Comparison of the clump number distribution in the inner 10 kpc of our fiducial MHD simulation at $t=960$ Myr with three different method. The inclination of a power-law of index $\alpha \sim -0.6$ (equivalent to a clump mass function $\mathrm{d}N/\mathrm{d}M_{\rm{clumps}} \propto M_{\rm{clumps}}^{-1.6}$) is indicated for comparison. The blue and orange curves shows the distribution of the clumps with \texttt{yt} \citep{Turk_2011} based on density and temperature criteria. The red curve indicates the clump distribution using Scipy's \texttt{ndimage} \citep{Virtanen_2020}, and is comparable to the method used in \citep{Gronke_2022} and \cite{Das_2024}.}
    \label{fig:ClumpFinderAlgo}
\end{figure}

As stated in Sect.~\ref{section:morphology}, the clump mass distribution around filaments in our MHD simulations is found to be characterized by a power law ranging from $\sim 10^5$ to $\sim10^8 \, \rm{M}_\odot$, with a power-law index $\alpha \sim - 1.6$. While broadly consistent with result from other studies, it is slightly shallower than the -2 slope frequently reported in previous studies \citep[e.g.]{Gronke_2022,Fieding_2023,Tan_2024,Das_2024,Augustin_2025}. In particular, the clump-finding algorithm used in \citet{Gronke_2022} and \citet{Das_2024} is different than ours, as it uses Scipy's \texttt{ndimage} \citep{Virtanen_2020} library to extract temperature contours, while our analysis is performed with \texttt{yt} clump-finding algorithm which performs a hierarchical search for structures recursively between two density extrema, similar to the analysis performed in \citet{Yuan_2014b}. To assess the sensitivity of our results on the choice of the clump-finding algorithm, we present in Fig.~\ref{fig:ClumpFinderAlgo} the clump distribution for the inner 10 kpc of our simulated box for our fiducial MHD run. The blue and orange curves are obtained with the \texttt{yt} clump finder. For the blue one, the standard hierarchical search of disconnected structures is performed such as described in Sect.~\ref{section:morphology}. For the orange curve, the selection is performed with \texttt{yt} based on a temperature criteria, such that clumps are defined as disconnected structures with a $T=10^5$ K contour. This second method is theoretically identical to the third one which is comparable to the method used in \citep{Gronke_2022} and \cite{Das_2024}. Here, the gas of temperature $T \leq 10^5$ K is isolated and Scipy's \texttt{ndimage} \citep{Virtanen_2020} is used to find the clumps. As visible, the second and last method are in a nearly perfect agreement, while the number of clumps is a factor of $\sim 2$ larger using the method employed for the analysis in this paper. This is due to the fact that clumps are not defined the same way between these two methods. Using a density criteria and a hierarchical arrangement of clumps allow to distinguish structures which are not distinguished by the \texttt{scipy} method. The slope of the distribution is consistent across methods.

\end{appendix}
\end{document}